\documentclass{emulateapj}
\newcommand{\feoh}{[{\rm Fe} / {\rm H}]}
\newcommand{\msun}{\, M_\odot}
\newcommand{\Zsun}{\, X_{\rm Fe, \odot}}
\newcommand{\mmd}{M_{\rm md}}
\newcommand{\dm}{\Delta_M}

\newcommand{\sfe}{c_\star}
\newcommand{\mh}{\, \mathchar`- \,}

\def\abra#1#2{[{\rm #1}/ {\rm #2}]}
\shorttitle{Formation History of Halo Stars}
\shortauthors{Komiya et al.}

\begin{document}
\title{Formation History of Metal-Poor Halo Stars with Hierarchical Model\\ and the Effect of ISM accretion on the Most Metal-Poor Stars}

\author{Yutaka Komiya\altaffilmark{1,2}, Asao Habe\altaffilmark{3}, Takuma Suda\altaffilmark{3,4}, Masayuki Y. Fujimoto\altaffilmark{3}}
\altaffiltext{1}{Astronomical Data Center, National Astronomical Observatory of Japan, Osawa, 181-8588, Japan}
\altaffiltext{2}{Astronomical Institute, Tohoku University, Sendai, Miyagi 980-8578, Japan}
\altaffiltext{3}{Department of Cosmoscience, Hokkaido University, Sapporo, Hokkaido 060-0810, Japan}
\altaffiltext{4}{Astrophysics Group, EPSAM, Keele University, Keele, Staffordshire ST5 5BG, UK}

\begin{abstract} 

We investigate the star formation and chemical evolution in the early universe by considering the merging history of the Galaxy in the $\Lambda$ cold dark matter scenario according to the extended Press-Schechter theory.   
    We give some possible constraints from comparisons with observation of extremely metal-poor (EMP) stars, made available by the recent large-scaled surveys and by the follow-up high-resolution spectroscopy. 
  We demonstrate that (1) The hierarchical structure formation can explain the characteristics of the observed metallicity distribution function (MDF) including a break around $\feoh=-4$. (2) A high mass initial mass function (IMF) of peak mass $\sim 10 \msun$ with the contribution of binaries, derived from the statistics of carbon enhanced EMP stars (Komiya et al. 2007, \apj, 658, 367), predicts the frequency of low-mass survivors consistent with the number of EMP stars observed for $-4 \lesssim \feoh \lesssim -2.5$.  
   (3) The stars formed from primordial gas before the first supernova (SN) explosions in their host mini-halos are assigned to the hyper metal-poor (HMP) stars with $\feoh \sim -5$. 
   (4) There is no indication of significant changes in the IMF and the binary contribution at metallicity  $-4\gtrsim \feoh \gtrsim -2.5$, or even larger as long as the field stars of Galactic halo are concerned..  

We further study the effects of the surface pollution through the accretion of interstellar matter (ISM) along the chemical and dynamical evolution of the Galaxy for low-mass population III and EMP survivors.   
   Because of shallower potential of smaller halos, the accretion of ISM in the mini-halos in which these stars were born dominates the surface metal pollution. 
  This can account for the surface iron abundances as observed for the HMP stars if the cooling and concentration of gas in their birth mini-halos is taken into account.  

We also study the feedback effect from the very massive population III stars. 
   The metal pre-pollution by pair-instability SNe (PISNe) is shown to be compatible with the observed lack of their nucleosynthetic signatures when some positive feedback on gas cooling works and changes IMF from being very massive to being high mass. 
   

\end{abstract}

\section{Introduction}\label{introS}    

In the early universe, the first stars should be formed from the primordial matter without metal elements
\citep[e.g.][and references therein]{Barkana01, Bromm04}.  
   These stars formed without metals are referred to as Population III stars. 
Despite longstanding efforts, however, direct proof of their existence has been elusive.  
   The first stars illuminate the universe and enrich surrounding matter with metals and the subsequent generations of stars with low metallicity should be formed under their influence.  
   At the early stages of formation and evolution of galaxies, the feedback from stars is thought to play an important role. 
   Since the era of the first stars and galaxy formation has occurred in the Dark Age of the universe, however, it is difficult to investigate them by direct observation with current instruments. 
   One promising observational probe into this era is the extremely metal-poor (EMP) stars, which were stars formed in the early stages of the Galactic chemical evolution.  
   The observations of EMP stars are expected to be connected with the first stars and Galaxy formation.  

Over the past decade, thousands of very metal-poor stars below $\feoh \simeq -2$ have been identified in the Galactic halo by two large-scaled surveys (HK survey, Beers et al.\ 1992: Hamburg/ESO [HES] survey, Christlieb et al.\ 2001).   
   The follow-up observations with high-dispersion spectroscopy (HDS) using large telescopes yield details of hundreds of these star's surface abundances \citep[e.g., see the review by][]{Beers05a}.  
   Thanks to these observations, now we are able to discuss statistical features of these stars \citep[see, e.g., SAGA database;][]{Suda08, Suda09}. 
   It is particularly worth noting that the most iron-deficient objects known to date are identified among the HES sample with much smaller metallicity than other EMP stars, i.e., with $\feoh=-5.6$, $-5.4$ and $-4.8$ (HE1327-2326, Frebel et al.\ 2005: HE0107-5240, Christlieb et al.\ 2002: HE0557-4840, Norris et al.\ 2007). 
   We refer to these three stars as hyper metal-poor (HMP) stars, distinct from EMP stars, defined as stars with $\feoh \lesssim-2.5$.  

The EMP stars display some features distinct from the Population~II stars and Pop.~III stars, as summarized in Section~\ref{EMPs}. 
   In previous studies \citep[][, hereafter Paper I and Paper II, respectively]{Komiya07,Komiya09}
   we propose that they compose a peculiar stellar population formed with a high-mass initial mass function (IMF).  
   We also argue that most of observed EMP stars were formed as the secondary members in binaries.   

From the observations of EMP stars in the Galactic halo, the scarcity of stars below $\feoh \simeq -4$ has emerged and has been investigated from the perspective of chemical evolution \citep[]{Prantzos03, Karlsson05}.  
    Among $\sim150$ stars with $\feoh<-3$ with HDS observations, only three HMP stars are discovered below $\feoh\leqq-4.8$ and no stars are identified between $-4.8<\feoh<-4$, although a few stars are found slightly below the $\feoh=-4$. 
   In Paper II, we propose a new scenario for this scarcity that the stars are divided into two groups according to whether they were formed before and after the parent mini-halos were polluted by their own first supernovae (SNe). 
HMP stars belong to the former group while the stars of the latter group constitute EMP stars. 
   We summarize this scenario in Section~\ref{cutoffS}.  

There have been investigations on the origins of the metal elements in the HMP stars.   
   All of three HMP stars, known to date, show large carbon enhancement.  
   \citet{Suda04} propose a binary scenario to assert that HMP stars can be the survivors of the low-mass Pop.~III stars without pristine metal elements.  
   In this scenario, their surface metal elements originate from afterbirth pollution by the accretion of interstellar matter (ISM) and by the wind accretion of envelope matter ejected by the binary companions. 
   \citet{Umeda03} and \citet{Iwamoto05} propose that HMP stars are formed of the gas, enriched with metals by the ejecta of peculiar faint Pop.~III SN.  
   \citet{Limongi03} propose that a combination with two Pop.~III SNe can reproduce the abundance pattern peculiar to HMP stars. 
   \citet{Maynet06} suggest that light metal elements on HMP stars are brought by the stellar wind from rapidly rotating Pop.~III massive stars.    
   For the binary scenario, \citet{Nishimura09} investigate the AGB nucleosynthesis under the dearth of pristine metals to show that the abundance patterns of light metal elements from C through Al can be reproduced in terms of the accretion of envelope matter from AGB companion in binaries.  
   In this case, the origin of the iron group elements remains to be discussed because binary mass transfer or stellar wind cannot provide these elements.  
   As for the SN scenario, there remains a problem of forming the next generation of stars with an appropriate mixture of SN ejecta and ISM, and in particular, for faint SNe with the peculiar abundances, the stars have to be formed with the efficient use of a relatively small amount of SN ejecta. 
   It is necessary, therefore, to discuss the origin of iron on the surface of HMP stars in the proper context of structure formation of the Galaxy.   

From the theoretical viewpoint, many calculations of the formation of first stars in the early universe have been carried out \citep[e.g.][]{Bromm99, Abel02, Omukai03, Tan04, O'Shea07, Yoshida08}.   
   These studies suggest that one very massive star with $M \gtrsim 100\msun$ may be formed in the primordial mini-halos with mass $10^6 \msun$, although typical masses or the mass range of the first stars are yet to be properly decided, especially in relation to the role of angular momentum in the fragmentation \citep{Clark08,Machida08}.  
   On the other hand, the evolution of Pop.~III and EMP stars have been investigated rather well, and in particular, it is shown that very massive stars with $M \sim 200\msun$ explode as pair-instability SNe (PISNe) with huge explosion energy and a large amount of metal yield \citep{Umeda02, Heger02}.  
   The effects of these energetic SNe on the subsequent star formation and galactic chemical evolution are not yet understood well, however \citep[e.g.][]{Machida05, Kitayama05, Greif08}.  
   The conditions for the formation of the first low-mass star are subject to controversy as well.  
   Some studies propose that metal enrichment by massive stars accelerates the cooling of molecular clouds to lower the Jeans mass and to enable the low-mass star formation \citep{BrommL03, Omukai05}. 
   Other studies propose that ionization photons emitted from first stars accelerate the formation of ${\rm H}_2$ molecules as main coolant in the primordial gas \citep{Ricotti02, Yoshida08a}, which can trigger the first formation of low-mass stars.  
   There remains the possibility that low-mass stars can be formed as the first objects \citep{Nakamura01, Clark08}.  

In this paper, we focus on the origin and nature of HMP and EMP stars and discuss the first stars, the early chemical evolution, and the star formation history of the Galaxy in the context of the hierarchical structure formation scenario.  
   We have built the merger trees of the Galaxy using a semi-analytic method and follow the history of star formation and chemical enrichment along the trees \citep[][hereafter, Paper III]{Komiya09L}. 
   We register all the individual Pop.~III and EMP stars and trace their formation and evolution. 

Firstly, we derive the MDFs and compare them with those of the observational counterparts in the Galactic halo. 
   We adopt the high mass IMF, based on the comparisons of stellar models with observations, and take account of the binary contribution according to Paper I and Paper II.  
   We also illustrate the chemical enrichment history in the hierarchical structure formation scenario and discuss the dependence of MDF on the assumptions about feedback from first stars and star formation rate in the early universe.  
   In our study, we take account not only of the shape of the MDF but also of the total number of EMP and HMP stars.

Secondly, on the basis of this scenario, we study the effect of surface metal enrichment by accreting the ISM and attempt to predict the observational abundances of the first low-mass stars.  
   Unlike the Pop.~I and Pop.~II stars, the ISM accretion can affect the surface metal abundance of Pop.~III and EMP stars because of their very low pristine metal abundances \citep{Yoshii81, Iben83}. 
   Furthermore, in small halos, in which Pop~III and EMP stars are born, the accretion rates of ISM gas are thought to have been much higher as compared with the those in the present Galactic halo because of much shallower gravitational potential. 
   In order to estimate the ISM accretion rate and the change in the surface abundances of halo stars in the early universe, therefore, we should follow the merging history of halo as well as the star formation and chemical evolution.  
   There are some previous studies of the ISM accretion \citep{Iben83, Frebel09} onto Pop.~III stars but none of them take account of the dynamical and chemical evolution self-consistently.  
   In this paper, we present the calculations of ISM accretion taking into account the merging history and the metal enrichment history of the Galaxy to explore the dependence of accretion rates on the dynamics of gas and stars.  
   We demonstrate that the ISM accretion provides a reasonable interpretation of the observed iron abundance of HMP stars.  

Thirdly we study possible metal pre-pollution by energetic SNe and draw observational constraints on the first stars and feedback from them.  
   If PISNe or hypernovae (HNe) with the explosion energy and iron yield much larger than those of normal core collapse SNe \citep{Heger02, Umeda02} occur in the low-mass halos, the blast wave may sweep out the gas from their host halos \citep{Bromm03, Kitayama05, Machida05}.  
   As a result, their metal ejecta is blown off from the host halos to pollute the inter-halo gas.   
   We calculate the MDFs with such effect due to PISNe according to the hierarchical chemical evolution program and compare them with the observed HMP and EMP stars.  

A semi-analytic approach to hierarchical structure formation has been adopted by \citet{Tumlinson06} and by \citet{Salvadori06}, who have constructed the merger trees of the Galaxy and investigated inhomogeneous Galactic chemical evolution models. 
   \citet{Tumlinson07L} considered an IMF dependent on the temperature of the cosmic microwave background, and discussed the carbon-rich EMP stars also by taking into account the contribution of binaries.  
   However, he did not relate hierarchical nature to the scarcity of the HMP stars.  
   We address the difference between the HMP and EMP stars in the context of hierarchical galaxy formation. 
   Additionally, we discuss the total number of EMP stars as well as shape of the MDF. 
   Furthermore, none of these studies is capable of explaining the origin of the extraordinarily low iron abundance of HMP stars.  
   We examine two possible origins, i.e., the surface pollution by ISM accretion and the pre-pollution of intra-Galactic matter (IGM) by halo blown off, for the small amount of iron in HMP stars.  
   
The paper is organized as follows. 
   We review the observations and elaborate our scenario in the next section and describe the computational method of star formation and evolution of Galaxy, in Section~\ref{numericS}.    
   The following three sections are devoted to presenting the results of our computations for the halo formation models, the surface pollution due to the ISM accretion, and the effects of pre-pollution of mini-halos by energetic PISNe, respectively. 
   We discuss the formation of the first stars and HMP stars in Section~\ref{discussion} and a summary of the conclusion follows in Section~\ref{summary}.  

\section{Observation and Scenario Preliminaries}

In this paper, we use the following terminology for the classification of metal-poor stars.  
\hfill\break
\vskip 3pt
   {\sl EMP stars:}  Stars with metallicity $-\infty < \feoh \lesssim -2.5$ different from those with of $-4 \feoh<-3$ in the literature like \citet{Beers05a} because the stars with $\feoh \lesssim -2.5$ show some theoretical and observational peculiarities, distinct from the stars of larger metallicity, as described in Section~\ref{EMPs}.  
   Stars formed with $Z=0$ are classified into the other population, Pop.~III stars.  
\hfill\break
\vskip 3pt
   {\sl EMP population:} Population of stars with metallicity $-\infty < \feoh <-2.5$.  
   It is the mother population of low-mass EMP survivors, and involves the stars with mass larger than $\sim 0.8\msun$ that have ended the nuclear burning stages.  
\hfill\break
\vskip 3pt
   {\it EMP survivors:} EMP stars with nuclear burning still going on.  
   They are the low-mass members of the EMP population with mass $M \lesssim 0.8\msun$. 
   Most of these should be the secondary companions of binaries under our assumed high-mass IMF (see \S\ref{EMPs} and Paper~I).  
   Among the EMP stars born in the early universe, it is only EMP survivors that are currently observed in the Galactic halo.  
\hfill\break
\vskip 3pt
   {\it HMP stars:} Stars observed with metallcity $\feoh<-4.5$.  Currently, three stars are assigned to this category.  
\hfill\break
\vskip 3pt
   {\it Pop.~III stars:}  Stars formed in primordial gas totally devoid of metals in their interior.  
   In addition to the first stars, some may be formed under the radiative feedback effect from the first massive stars.  
   Low-mass Pop.~III stars are, if formed, still shining as with the nuclear burning today, and are referred as {\it ``Pop.~III survivors"}.  
   Pop.~III stars may suffer from the surface pollution by ISM accretion after their birth and appear enriched with finite metal abundances on their surface today.   
   We refer to these stars, which are formed as Pop.~III stars but no longer $Z=0$ by surface pollution, as {\it ``polluted Pop.~III stars"}. 
   In section \ref{accS}, we discuss the possibility that HMP stars are polluted Pop.~III stars.  

\subsection{Peculiarities of Stars with $\feoh\lesssim-2.5$}\label{EMPs}


EMP stars are formed of gas that contains small but finite abundances of metal elements, synthesized in stars of previous generations, and are distinct from Pop.~III stars. 
   These stars are also distinguished from Pop.~II stars by peculiar features, as summarized below.  

Theoretically, the evolution of low- and intermediate-mass stars of metallicity $\feoh\lesssim-2.5$ differs from that of Pop.~II stars \citep{Fujimoto90,Fujimoto00}. 
   A mixing mechanism called He-flash driven deep-mixing works in the EMP stars at AGB phase. 
   Carbon and the {\it s}-process elements are dredged up to their surface by this mechanism. 

Observationally, some peculiarities have been pointed out, too. 
   It is known that the large portion ($20-25\%$) of EMP stars show carbon enhancement \citep{Rossi99,Christlieb03} in contrast with a small fraction ($\sim 1\%$) of CH stars among Pop.~II stars.  
   In addition, some carbon-enhanced EMP stars stand without s-process element enhancement \citep{Aoki07}, while no CH stars do. 
   The number of EMP stars is very small compared to Pop.~I and Pop.~II stars, which may be referred to as the G-dwarf problem in the Galactic halo. 
   All globular clusters have the metallicity $\feoh > -2.5$ and the local dwarf galaxies show the metallicity distribution of stars that steeply decreases around $\feoh\sim-2.5$ with decreasing metallicity \citep{Helmi06}. 

In Paper~I, we show that the IMF of the EMP population is different from Pop.~II stars, based on the observed statistics of the CEMP survivors. 
   Only massive IMF with medium mass $M_{\rm{md}}\sim10\msun$ can form simultaneously a large fraction of CEMP stars and the number ratio between the CEMP stars with and without the enrichment of the {\it s}-process elements as observed among the EMP stars. 
   This involves the consequence that only a small fraction of EMP population stars can be observed today as EMP survivors and that most survivors are the secondary components of binaries.  
   The effects of high-mass IMF on the chemical enrichment are discussed by a simple one-zone model in Paper II and shown to be consistent with the observed small number of EMP survivors.  
   This means that the formation process of EMP stars differs from metal-richer Pop.~II stars. 

\subsection{Metallicity Gap and Hierarchical Chemical Evolution Scenario}\label{cutoffS}

Scarcity of stars with $\feoh\lesssim -4$ and the existence of HMP stars have drawn wide attention, but are yet to be fully clarified. 
   \citet{Prantzos03} discusses the low-metallicity tail of the MDF, suggesting that the problem can be alleviated by introducing an early gas infall. 
   \citet{Karlsson05} points out that such MDF break can be interpreted as a result of stochastic metal pollution process.  
   \citet{Karlsson06} proposes that HMP stars are formed of the matter, pre-polluted by SNe ejecta and polluted by carbon-rich stellar wind following the suppression of star formation after the first SNe.  
   Since the Pop.~III stars are thought to be massive, some studies assume the change of IMF into that which enables the formation of low-mass stars somewhere below $\feoh \simeq -4$ \citep[e.g.,][]{Tumlinson06, Salvadori06}, but some recent studies show low-mass stars can be formed in a more metal poor environment\citep{Omukai05, Schneider06}. 

We have proposed a new scenario to explain this break from the viewpoint of the structure formation process in Paper~II. 
   According to the current $\Lambda$ cold dark matter (CDM) model, galaxies are formed hierarchically;  
   Low-mass structures are first formed, and then, merge with each other and/or accrete matter to form galaxies like the Milky Way. 
   Stellar halos are thought to be formed through the accretion of many small halos \citep{Searle78}.  
   In early stages of the Galaxy formation, star forming gas is incorporated into many small halos. 
   The mass of the first star forming mini-halos is thought to be $\sim10^6 \msun$ with a baryonic content of $\sim 2 \times 10^5 \msun$\citep{Tegmark97, Nishi99}. 
   When a massive first star explodes as core collapse SN, it enriches the host mini-halo with iron ejecta of $\sim 0.07 \msun$ up to 
\begin{equation}
\feoh = \log \left( \frac{0.07}{2\times 10^5} \right) -\log \Zsun \sim -3.5, 
\end{equation} 
   on average, where $\Zsun$ is the solar iron abundance. 
   We refer to this event as {\it ``first pollution"}, which sets a lower limit to the metallicity of the second and subsequent generations of stars.  
   In some mini-halos, the metallicity decreases by the accretion of gas and/or by merging with mini-halo with pristine abundance after the first pollution, which enables the formation of stars with smaller metallicity of $\feoh \sim -4$. 

In this scenario, stars of metallicity much lower than $\feoh \simeq -4$ such as HMP stars must be assigned to the survivors of stars formed prior to the first pollution. 
   They are the {\it ``first low-mass stars"} and their surface metal elements are provided either by the surface pollution after their formation and/or by pre-pollution of IGM before the formation of the host halo.  
   We explore this scenario taking into account the merging history of the Galaxy, and investigate the current appearance of the first low-mass stars in Paper III. 
   In this paper, we investigate some models of the metal abundance of HMP stars quantitatively and discuss their formation in detail.  

\section{Models and Computation Method}\label{numericS} 

\subsection{Merging History --- the Extended Press-Schechter Theory
} \label{merging_history}

First, we build merger trees of our Galaxy using the extended Press-Schechter (EPS) approach \citep{Bond91, LC93}.  
   The Press-Schechter theory \citep{PS74} is an analytical formalism that can draw halo mass distribution in the universe.  
The extended Press-Schechter theory gives the distribution of mass, $M_1$, of building blocks at redshift $z_1$ that yields a halo of a prescribed mass $M_2$ at $z_2$, and provides a method of tracing back through the formation history of halos.  
   The mass distribution of such building blocks is given in the following equation,   
\begin{eqnarray}
f(S_1, \omega_1, S_2, \omega_2 )dS_1
 &=& \frac{\omega_1-\omega_2 }{(2\pi )^{1/2}(S_1-S_2)^{3/2}} \nonumber \\
&& \times \exp \left[ -\frac{ ( \omega _1- \omega _2 )^2 }{ 2( S_1-S_2 ) } \right] dS_1,
\end{eqnarray}
   where $\omega_{\{1,2\}} = \omega(z_{\{1,2\}})$ is critical density to collapse for linear growth theory of overdensity at $z_{\{1,2\}}$ and $S_{\{1,2\}} = S (M_{h, \{1,2\}})$ is the variance of the density field smoothed on mass scale $M_{h, \{1,2\}}$.  
   We use this equation to determine the distribution of mass $M_{h, 1}$ of the halos at $z_1$ that eventually form the halos of mass $M_{h, 2}$ at $z_2$.

To calculate the merger tree from the EPS theory, we adopt the N-fold accretion method constructed in \citet{SK99}.  
   In this method, a merger tree starting from the halo of a prescribed mass branches and accretes material as we step back to higher redshift. 

The low mass limit of mini-halos in which stars are formed is a critical parameter to plant the merger tree and to determine the start of chemical evolution. 
   When a halo and gas cloud collapse, the cloud virializes and is heated to the virial temperature of $T_{\rm vir}$. 
   Dependence on mass and redshift at $z\gg1$ is described as
\begin{equation}
T_{\rm vir} \propto M_h^{2/3}(1+z).
\end{equation} 
   It is shown that for $T_{\rm vir}(M_{h}, z) \gtrsim 10^3{\rm K}$, $\rm{H_2}$ molecules form and function effectively as a coolant and the primordial gas clouds can cool in Hubble time \citep{Tegmark97, Nishi99}. 
   This sets the lower limit, $M_{h, l}$, to the total (baryon + dark matter) mass of merger tree, which is written in the form; 
\begin{eqnarray}
M_{h, l} &=& 1.14 \times 10^6 \msun \left( \frac{H_0}{100 \hbox{ km s}^{-1} \hbox{ Mpc}^{-1} } \right)^{-1} \nonumber \\
&&\times \left( \frac{1+z}{10}  \frac{\mu}{0.6} \right) ^{-3/2}  \left( \frac{\Omega_m}{\Omega^z_m} \frac{\Delta_c}{18\pi^2}  \right)^{-1/2}.
\label{M_hl}
\end{eqnarray} 
    Here $\mu$ is the mean molecular weight of gas:
    $H_0$ is the Hubble parameter:
    $\Omega_m^z = \Omega_m/(\Omega_m +\Omega_k/(1+z)^2 +\Omega_\Lambda /(1+z)^3) $: 
    and $\Delta_c$ is the overdensity of virialized objects relative to the critical density. 
   As $z$ increases, $M_{h, l}$ decreases, i.e., a halo resolves into mini-halos having smaller masses. 
   The low-mass halos at tips of the branches of the merger tree are regarded as the hosts of the first stars. 

\begin{figure}
\plotone{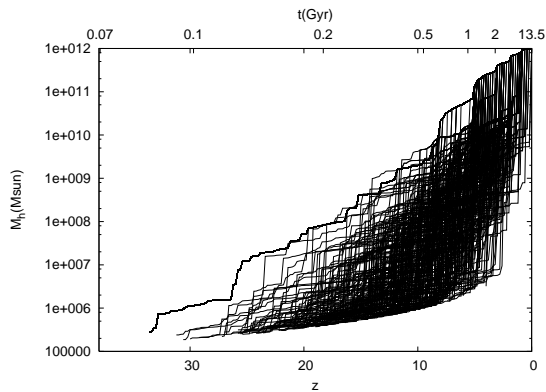}
\caption{
   An example of merger trees for a halo of mass $10^{12}\msun$. 
   The variations of total (dark matter plus baryon) mass, $M_h$, of mini halos are plotted against the redshift, $z$, (bottom) or the age, $t$, of the Universe (top) for 1000 branches randomly selected among the total of $\sim 2 \times 10^5$.  }\label{tree}
\end{figure}

Figure~\ref{tree} illustrates the merger tree of a halo with mass similar to the Galaxy ($10^{12}\msun$) in $\Lambda$CDM universe ($\Omega_M=1-\Omega_\Lambda=0.3, \Omega_b=0.045, h=0.7, \sigma_8=0.9$). 
    The number of branches amounts to $\sim 2 \times 10^5$ but here we plot only one thousandth of total branches.  
    The first mini-halo is formed at $z \gtrsim 30$ with mass lower than $10^6\msun$.  
    For a merger tree with cosmological parameters by WMAP 5 years, the number of branches is $\sim10\%$ smaller and their typical age of formation is $\sim 20\%$ later, but our conclusion is almost the same for this model. 
    Most mini-halos are formed between $z=20$ and $z = 5$, which corresponds to the ages of $t \sim 0.15$ Gyr and $\sim1$ Gyr from Big Bang, respectively, while some young halos are formed at later ages.  

\subsection{Star Formation and Chemical Evolution}

We first summarize the assumptions about star formation of EMP and Pop.~II stars and about chemical evolution.  
   As for the formation of first-generation stars including Pop.~III stars and their feedback effects, we test various models as described in the next subsection.  

\begin{itemize} 
\item
Star formation rate (SFR), $\psi$, in the halo with gas mass, $M_{\rm gas}$, is assumed to take the form of $\psi = \sfe M_{\rm gas}$ after the first SNe in host halos, 
   where $\sfe$ is the star formation efficiency (SFE).  
  The value of $\sfe$ is set at $10^{-10} \hbox{ yr}^{-1}$ for our fiducial model, which means that gas is converged to stars in 10 Gyr.  
   We also adopt that $\sfe=10^{-9} and 10^{-11} \hbox{ yr}^{-1}$ to see the dependence on SFE. 
   For the fiducial model, SFR becomes $\psi = 10^{-4} \msun \hbox{ yr}^{-1}$ for halos with gas mass $10^6\msun$.  
   We register all the individual stars for the EMP population and follow their evolution. 

\item
 The mass of each star is specified randomly according to the statistical weight with the initial mass function (IMF) at its birth.  
   We assume a lognormal IMF.  
\begin{equation}
\xi (m)\propto \frac{1}{m} \exp \left[ -\frac{ \{ \log (m/ \mmd) \} ^2}{2\times \dm^2} \right] .
\label{eq:IMF-lognormal}
\end{equation}
For EMP population stars, we assume high-mass IMF, derived by \citet{Komiya07}, i.e., lognormal IMF with the medium mass $\mmd = 10\msun $ and the dispersion $\dm = 0.4$.  
   For Pop.~II stars with the metallicity of $ \feoh \geq -2.5$, we calculate two cases; 
   the same high-mass IMF as the EMP stars, and the low-mass IMF, base on the observations of the Galactic halo stars of $\feoh > -1.5$ \citep{Chabrier03}, the latter of which is given by a lognormal mass function in eq.~(\ref{eq:IMF-lognormal}) with $\mmd=0.22 \msun$ and $\dm = 0.33$ for $m \le 0.8 \msun$ and the Salpeter mass function for $m > 0.8 \msun$, respectively. 

\item
 Half of all stellar systems are taken to be binaries. The distribution, $n (q)$, of mass ratio, $q \ (\equiv m_2/m_1)$, between the secondary and primary stars of binary systems is assumed to obey a flat distribution, i.e., 
\begin{equation}
n(q)=1. 
\end{equation}
\item
 Massive stars with $m=10-40\msun$ explode as core collapse SNe after $\tau =10^7$ yrs from their birth and eject iron yield of mass $M_{\rm Fe} = 0.07 \msun$ on average. 
 Very massive stars with $m=140-260\msun$ explode as PISN and eject iron yield of mass $M_{\rm Fe} = 10 \msun$ in the lifetime of $\tau = 2 \times 10^6$ yrs
\item
 SNe ejecta is assumed to spread instantaneously and homogeneously in their host halos. 

\item
In the fiducial model, gas is regarded as retained within the host mini-halo.   
   We also compute models with gas ejection from the halos.   
   All gas and metal are assumed to be blown off the host mini-halo when the kinetic energy of a shell formed by supernova is larger than binding energy $E_{bin}$ of the gas in host mini-halo, i.e., $\epsilon_k E_{SN}>E_{bin}$, where $E_{SN}$ is the energy ejected by supernova and $\epsilon_k$ is the fraction of kinetic energy.  
   We take $E_{SN}=10^{51}erg$ for type~II SN and $E_{SN}=10^{52}erg$ for PISN, and $\epsilon_k=0.1$ for model with blowout.  
   For simplicity, ejected gas and metal are assumed to be mixed with gas throughout the Galaxy instantaneously and homogeneously.  


\item
Baryon fraction in primordial mini-halos is assumed to be the same as that of the average in the universe, $\Omega_b/\Omega_M$.  
   When gas and metals are blown off by energetic supernova, $M_{\rm gas}$ vanishes, but dark matter halo remains. 
   Afterward, blown-off mini-halos accrete dark matter and gas at the mass accretion rate described in the merger tree.  
   When gas mass reaches $M_{h,l} \Omega_b/\Omega_M$, the star formation starts again.  

\end{itemize}

Since we are interested in the global features of Galactic EMP stars, we deal only with iron as a typical indicator of metal-enrichment in the universe.  
   The detailed abundance pattern may depend on the models and yields of SN explosions and the dilution process of the ejecta in the halo gas.  
   These effects may be subject to large uncertainties. 

For metal pollution of IGM, we make simplified assumptions and discuss the overall trend. 
   In this paper, we apply a semi-analytic method and discuss the dependence on the feedback effects from massive stars. 
   Galactic wind from more massive galaxies also contributes to the IGM metal pollution but we neglect them. 
   As shown in Section~\ref{firstS}, gas blown off the mini-halos by primordial stars is the dominant metal source for HMP stars on which we are focused. 
   In actually, IGM metal enrichment is an inhomogeneous process and demands further study including effect of the wind originated from Pop.~II stars \citep{Tornatore07, Trenti10}.  
   
\subsection{Primordial Stars and Feedback}
   Stars formed without any feedback effect from previous stars are referred to as primordial stars. 
   These stars, born prior to EMP stars referred to as primordial stars, can be very massive and exert various influences on the star formation processes of subsequent generations.  
   In Table~\ref{Tmodels}, we summarize models parameters for this feedback. 
   We present the most simplified model without any feedback from massive stars as fiducial models.  
   In this section, we describe other models with feedback effects and model parameters.  

\begin{table*}
\begin{center}
\caption{Model Parameters}
\label{Tmodels}

\begin{tabular}{l|cllllc}
\tableline\tableline
name  & photodissociation & cooling & $\sfe (yr^{-1})$ & $M_{\rm md,p} (\msun)$ & $\epsilon_k$ & accretion\\
\tableline
F0 & no	& & $10^{-10}$	& 10 & 0 & \\
Fs \tablenotemark{a} & no	& & $10^{-10}$	& 10 & 0 & \\
F-9 & local\tablenotemark{b}	& & $10^{-9}$	& 10 & 0 & \\
F-10 & local	& & $10^{-10}$	& 10 & 0 & C\\
F-11 & local	& & $10^{-11}$	& 10 & 0 & C\\
F100 & local	& & $10^{-10}$	& 100 & 0 & \\
\\
FA & local	& & $10^{-11}$	& 10 & 0 & A \\
FB & local	& & $10^{-11}$	& 10 & 0 & B \\
FD & local	& & $10^{-11}$	& 10 & 0 & D \\
\\
Vno & global\tablenotemark{c}	& $T_{\rm vir}>10^4$K	& $10^{-11}$@& 10@& 0 & \\
V & global	& $T_{\rm vir}>10^4$K	& $10^{-11}$	& 200	& 0.1 & \\
Pno & global	& photoionization	& $10^{-11}$	& 100	& 0 & \\
P & global	& photoionization	& $10^{-11}$	& 200	& 0.1 & \\
P10 & global	& photoionization	& $10^{-11}$	& 10	& 0.1 & \\
P100 & global	& photoionization	& $10^{-11}$	& 100	& 0.1 & \\
M & global	& $\feoh>-6$	& $10^{-11}$	& 200	& 0.1 & \\
M-4 & global	& $\feoh>-4$	& $10^{-11}$	& 200	& 0.1 & \\
M-10 & global	& $\feoh>-6$	& $10^{-10}$	& 200	& 0.1 & \\
\tableline
\end{tabular}

\tablecomments{Model parameters investigated in this paper. 
Column ``photodissociation'' denotes the assumption about the suppression of star formation by Lyman Werner photon (Sec.~\ref{IMFass}). 
Column ``cooling'' denotes the criterion for the second star formation with positive feedback effects and the change of IMF (Sec.~\ref{feedass}). 
$\sfe$ is the star formation efficiency, $M_{\rm md, p}$ is the medium mass of the primordial stars, and $\epsilon_k$ is the proportion of SN energy, made up by the kinematic energy for gas blown off the host mini-halos.  
Last column ``accretion'' denotes the model for the surface pollution of Pop.~III and EMP stars (Sec.~\ref{AccSec}). 
See text for detail. }
\tablenotetext{a}{For stars with $\feoh>-2.5$, the low mass IMF \citep{Chabrier03} is assumed. }
\tablenotetext{b}{In the mini-halos, star formation is suppressed between the formation and SN explosion of the first massive stars in these halos.}
\tablenotetext{c}{Lyman Werner background suppresses star formation of halos with $T_{\rm vir} < 10^{-4}$ K at $z < 20$}

\end{center}
\end{table*}

\subsubsection{Initial Mass Function}\label{IMFass}

    We assume the same high mass IMF with $\mmd=10$ both for EMP stars and for Pop.~III stars of the fiducial model, although typical mass of primordial stars can be much higher. 
    Previous numerical simulations of the first star formation show that, in the first star forming clouds, only one massive dense clump with mass $M \gtrsim 100 \msun$ is formed \citep[e.g.][]{Bromm99, Abel02}.   
   Typical mass, $M_{\rm md,p}$, of primordial stars is one free parameter in our model. 
   We calculate models with $M_{\rm md,p}=10,100,200\msun$. 
   For all the cases, we assume a lognormal IMF with $\dm=0.4$.  
   For the model with $M_{\rm md,p}=200\msun$, most of primordial stars become PISNe.  
   For the model with  $M_{\rm md,p}=100\msun$, stars with typical mass end their lives without iron ejection.  
   Recent simulations show that binary can be formed for first stars\citep{Turk09, Stacy09}.  
   In this study, the binary fraction and mass ratio distribution of primordial stars are assumed to be the same as those of EMP stars.  

\subsubsection{Negative Feedback by Lyman-Werner photon}\label{feedass}

For the Pop.~III star formation, $\rm{H_2}$ molecules work as coolant in mini-halos with $T_{\rm vir}>10^3$ K.  
   But once a massive star is formed in the primordial gas clouds, whole H$_2$ molecules are dissociated by radiation from the massive star to quench subsequent star formation \citep{Omukai99} until it explodes as SN. 
   In addition, the formation of low-mass stars may be promoted by SN explosions after the first pollution \citep{Machida05}. 
   Therefore, it is speculated that SFE was lower before the first pollution. 
   We calculate models with and without this local negative feedback.  
   For the model with local negative feedback, we assume that no stars are formed in the mini-halos during the lifetime of first stars.  

We also calculate models with global effect by Lyman Werner photons.  
   Lyman Werner photons can dissociate $\rm{H_2}$ molecules also in IGM and suppress star formation.  
We assume that Lyman Werner background from primordial stars becomes effective at z<20.  
   We assume that all mini-halos with $T_{\rm vir}>10^3{\rm K}$ can form primordial stars for $z>20$, but in the models with global negative feedback, mini-halos can form stars at $z < 20$ only when additional coolant is available. 
   For more massive halos, on the other hand, LW background does not affect the star formation because of self-shielding. 
   Additionally, for all the mini-halos with $T_{\rm vir}>10^4K$, we assume that stars are formed since gas is ionized by accretion shock and cools efficiently \citep{Uehara00}. 

The assumption on the LW photons is specified in column ``Photodissociation'' of Table~\ref{Tmodels}.  
   We calculate three cases of models without negative feedback, models with feedback only in the host halo, and models with feedback on the IGM. 
   In the table, these cases are referred to as no, local and global, respectively. 

\subsubsection{Positive Feedback and Transition of the IMF}

Some possible positive feedback effects on the star formation from primordial stars are also discussed.  
   They can reduce the low mass limit of mini-halos for the star formation and also the typical mass of stars. 
   
One is photo-ionization.  
   Once gas is ionized, H$_2$ molecules form efficiently with residual free electrons as catalysts.   
   Further, HD molecules act as main coolant to form stars in the high-density cooled gas.  
   Since the first very massive stars emit a large amount of ionization photon to ionize the intergalactic medium, some halos with $T_{\rm vir} < 10^4$ K are subject to the effect of the photo-ionization before they collapse.  
   In the halos with gas once photo-ionized, matter also cools by H$_2$ line emissions effectively \citep{Ricotti02}.  
   Ionization can also reduce the typical mass of stars of the subsequent generations \citep{Uehara00, Yoshida08a}.  

The other possible catalyst is the metal elements and dust.  
   The first very massive stars explode as SNe to pollute the ambient gas with metals and dust.  
   Line emissions from metal elements such as carbon and oxygen and/or thermal emissions from dust become dominant coolants for the polluted gas.  
   Additionally, H$_2$ molecules are formed on the surface of dust and lower the typical mass of stars formed.   
   Observationally, the detection of the HMP stars with $\feoh < -5$ indicates that some low-mass stars are formed in the gas below $\feoh<-5$.  
   \citet{BrommL03} propose that carbon of abundance larger than $\abra{C}{H}>-3.5$ leads the cloud to fragment into smaller clumps.  
   \citet{Omukai05} argue that for gas with metallicity $\feoh \gtrsim -5$, the effective cooling due to H$_2$ formed on the dust surface enables the low-mass fragmentation while gas with $\feoh \lesssim  -6$ evolves like the primordial gas.  
   \citet{Schneider06} show that dust synthesized by the first supernovae lowers the Jeans mass and enables low mass star formation at $\abra{Z}{H} \gtrsim -6$. 

We investigate some models with the conditions for the formation of second generation stars under positive feedback.  
   The assumptions are given in column "cooling" in Table~\ref{Tmodels}.  
   The first model "$T_{\rm vir}>10^4$K" shows no positive feedback effect from stars. 
   For this model, stars are formed only in halo with $T_{\rm vir}>10^4$ K for $z<20$ because of negative feedback by LW photons.  
   Models with this assumptions are named V. 
   The second model "photoionization" shows the effect of photoionization of IGM.  
   In the mini-halos formed of pre-ionized IGM, gas can cool and stars are formed.  
   We treat the ionization of ambient gas by the massive stars in the following statistical manner.   
   One Pop.~III star emits ionizing photons and increases the mass, $M_{\rm ion}$, in the ionized region of IGM by $\Delta M_{\rm ion} = N_{\rm ion} f_{\rm esc} m_{\rm p}$, where $m_{\rm p}$ is the mass of proton. 
   We use the number of ionization photons, $N_{\rm ion}$, emitted by massive stars by \citet{Schaerer02} and the escape fraction, $f_{\rm esc}$, from host halo by \citet{Yoshida08a}. 
   New halos are formed randomly in the ionized or unionized region.  
   The probability that a new halo is pre-ionized in advance of formation is given by $M_{\rm ion}/(10^{12}\msun \Omega_b/ \Omega_M)$, in which stars are formed under the same condition as the fiducial model.  
   The third is the metal-pollution model.  
   In the halos, pre-polluted by metals ejected from the mini-halos, gas can cool and form stars.  
   We introduce the critical metallicity, $\feoh_{\rm cri}$ for the formation of the second generation stars, and test the models with $\feoh_{\rm cri }=-6$ and $\feoh_{\rm cri }=-4$.  

As stated above, it is thought that primordial stars are very massive and EMP stars are less massive than primordial ones. 
   There should be a switchover of IMF from the very massive ($\mmd=200\msun$) to the high-mass one ($\mmd=10\msun$) with the formation of low-mass secondary survivors in the binaries. 
   We assume that all stars formed with positive feedback obey the IMF with $\mmd=10\msun$.  
   The IMF of stars, formed in the halo with $T_{\rm vir}>10^4\rm K$, is also a high-mass one with $\mmd=10\msun$.

\subsection{Surface Pollution by ISM Accretion}\label{AccSec}

We assume the Bondi-Hoyle accretion rate, 
\begin{equation}\label{Bondi}
\dot{M} = 4\pi (GM)^2 \rho / (V_r^2+c_s^2)^{3/2}, 
\label{mdot}
\end{equation} 
   where $\rho$ and $c_s$ are the density and sound velocity of the ISM, $V_r$ is the velocity of the stars (or the barycenter of binaries) relative to the ambient gas, and $M$ is the stellar mass (or the total mass of binary).  
   The intermediate-mass stars with initial mass $M$ are assumed to become white dwarfs with $0.6 \msun$ after their lifetimes of $10^{10}\times (M/\msun)^{-3}$ yrs. 
   In our assumption, massive stars of $M \ge M_{\rm up} (= 8\msun)$ explode as SNe, and after that, the secondary stars are released from the binary systems and become single stars.  

   Accreted matter is mixed in the surface convective zone with mass $M_{\rm SCZ}$.  
   The surface abundance of polluted Pop.~III star then becomes 
\begin{equation}
\feoh_{\rm p} = \log \left[\int_{t_0}  X_{\rm Fe, ISM} (t) \dot{M}(t) dt / (M_{\rm SCZ} \Zsun ) \right],  
\end{equation}
where $t_0$ is the time when the star is born, 
   $X_{\rm Fe, ISM} (t) $ is the iron abundance of ISM in their host halo as a function of $t$, taken from our fiducial model with the radiative feedback and model with lower SFE (F-10 and F-11). 
   Since most observed objects have the mass $M \simeq 0.8 \msun$, we calculate the surface metallicity distribution of Pop.~III and EMP survivors with $0.8\msun$ by assuming the surface convection of mass $M_{\rm SCZ}=0.003\msun$ and $0.2\msun$ for dwarfs and giants, respectively \citep{Fujimoto95}. 

There are several factors that may affect the ISM accretion and should be considered in the hierarchical structure formation.  
   Since the accretion rate in eq.(\ref{Bondi}) is dependent on the dynamics of gas and stellar systems, therefore, it may change as halos merge and grow, and yet, their effects are not yet revealed. 
   We calculate the ISM accretion for four models with simplified assumptions on the dynamics and distribution of stars and gas in the mini-halos in the early universe. 
   We use models without gas blown off and pre-polluted in order to see only the effect of surface pollution to the most metal-poor stars.   

In the simplest model of Case A, we assume that stars always move with the virial velocity, i.e., $V_{\rm vir}$, in ISM of average density equal to the gas density, $\rho_{\rm vir}$, of virialized mini-halos.  
   Since the virial velocity depends on the halo mass [$V_{\rm vir} \propto M_{h}^{1/3}(1+z)^{1/2}$] and the density of virialized halos is $\sim 200$ times the mean density of the universe at the collapse time [$\rho_{\rm vir} \propto (1+ z )^{3}$], the accretion rate is higher in smaller halos collapsed in higher redshift, as given by  
\begin{eqnarray}\label{eq:mdot}
\dot{M} &=& 4\pi (GM)^2 \rho_{\rm vir} V_{\rm vir}^{-3} \nonumber \\
 &\simeq& 4.3 \times 10^{-15} \left(  \frac{M}{0.8\msun}\right)^2  \nonumber \\
&& \left( \frac{M_h}{10^6\msun}\right)^{-1}  \left( \frac{1+z}{10}\right) ^{3/2}  \msun/{\rm yr}.  
\end{eqnarray} 

In actuality, gas and stars are concentrated around the center in the mother clouds in which Pop.~III and EMP stars are formed so that the relative velocity is lower than $V_{\rm vir}$ and the gas density is higher than $\rho_{\rm vir}$. 
   In the primordial clouds, gas cools to $\sim 200$ K with $\rm{H_2}$ molecule as the main coolant and falls toward the center to form stars. 
   For the model of Case B, we take $V_r$ to be comparable with the sound velocity at this temperature. 
   For the density of ISM, we assume an isobaric contraction from the virial temperature to 200 K, 
\begin{equation}\label{eq:rho200}
 \rho=\rho_{\rm vir } \left({200\rm{K}}/{T_{\rm vir}}\right).
\end{equation} 
   As the density increases in inverse proportion to the temperature and the relative velocity decrease as $V^2 \propto T$ so that the accretion rate increases as $\dot{M} \propto \rho V^{-3} \propto T^{-2.5}$.  
   Since the masses of host halos of Pop.~III stars are equal or slightly larger than the low-mass limit for star formation, the temperature of virialized gas is $T \simeq T_{\rm vir}(M_{h,l}) =1000$ K.  
   Accordingly, the accretion rate may be enhanced by 
\begin{equation}
\label{eq:caseB}
\dot M \propto \left( 200{\rm K}/1000{\rm K} \right)^{-2.5} \sim 56 .  
\end{equation}
   After the host halo merges to a larger halo, the stars are assumed to move with the virial velocity, $V_{\rm vir}$, of the merged halo in the ISM of its virial gas density, $\rho_{\rm vir}$.  

In addition, we have to take into account the time delay between the merger of dark-matter halos and the change of motion of baryonic systems since the incorporation of over-density region of dark matter as given by merger trees does not imply the immediate viralization of the constituent systems.  
   The baryonic systems in the merged mini-halo orbit as satellites for some time, and the relative velocity of stars to ambient gas may not change immediately. 
  \citet{LC93} calculate the time for a satellite's orbit to decay by dynamical friction on the basis of the standard Chandrasekhar formula. 
   In the model of Case C, we consider this delay time, and assume that the relative velocity of stars to ISM remains the same as in the satellite baryonic systems for dynamical friction timescale.  
   Other parameters are taken to be the same as in the model of Case~B. 

Finally, since most EMP stars are formed as the secondary members of binary systems under our high-mass IMF, we have to consider the allotment of accreted matter onto the binary systems to each member in order to estimate the pollution of surface abundances. 
   It is true, however, that the gas dynamics of accreted flow in the binary systems is yet to be established \citep{Bate97,Ochi05}.    
   While we simply assume that a half of accreted ISM settles onto each member in the models of Cases A-C, we compute an additional model of Case D where the members of binary systems accrete mass independently at a rate given by eq.(\ref{mdot}) by taking the other parameters in the same way as in the model of Case C.

\subsection{Comparisons with Observations}

In our study, we consider not only the pattern of MDF but also the total number of EMP survivors observed in our Galaxy because of the large survey area and the significant efficiency of the identification of stars by the HES survey. 
   The effective survey area is $S = 6726 \hbox{ deg}^2$ \citep{Christlieb08}. Roughly $40\%$ of the candidates, selected by the objective-prism survey, have been examined by the spectroscopic follow-up observations with medium resolution \citep{Beers05}. 
   In addition, the large limiting magnitude ($B <17.5$) assures that almost all of the EMP giant stars present in the survey fields are identified.    
   Therefore it is expected that the ratio of the number of giants detected in the survey to the total number of giants in the Galaxy should be $S/4\pi$. 
   As for dwarf stars, only nearby stars can be observed because of lower luminosity, for which we adopt the observed number ratio of EMP dwarfs to giants ($=0.93$). 

For stars with $\feoh<-2$, we use MDF of stars observed in HES survey for comparison \citep{Beers05}.  
   As for the stars of metallicity below $\feoh = -3$, it is possible only through the follow-up observations with high dispersion spectroscopy (HDS) to distinguish their metallicity.  
   We use data compiled with SAGA database \citep{Suda08}, which collects all the abundance data of metal-poor stars with $\feoh \le -2.5$ obtained with the HDS follow-up and published up to date.  
   The follow-up HDS observations are reported only for 153 stars for $\feoh < -3$, detected in HES survey and HK survey, while HES surveys counts $\sim 200$ candidates of stars with $\feoh<-3$.  
   For $\feoh<-3$, we scale MDF of SAGA by factor $200/153$, since we may well regard the selection of target stars for the follow-up observations as being unbiased below the metallicity of $\feoh \simeq -3$.  
   In actuality, three HMP stars currently known were all identified by the follow-up observations of HES samples. 
   We may expect to have $3 \times (200/153) \simeq 4$ HMP stars among the stars in the fields observed by HES surveys. 

For stars with $\feoh>-2$, we use kinematically selected sample of halo stars by \citet{Ryan91} because the sample of HES survey is biased to low-metallicity stars.  
   We scaled their MDF to match the MDF by HES survey at $\feoh=-2$.  
   Observed MDF is plotted in Figure~\ref{MDF} by solid line.   
   Recently, \citet{Schorck08} assess the selection bias of HES survey and estimate the corrected MDF. 
   Their result shows a steep increase in the number of stars above $\feoh > -2.5$ as plotted in Fig.~\ref{MDF} by dashed line, which is inconsistent with MDF by \citet{Ryan91}, waiting for further investigations.  

\section{Result 1: MDF and Formation History of EMP stars}\label{resultS}

In this section, we show results of our fiducial model and the models labeled with F without global feedback effects.  

\subsection{Metallicity distribution function}
\begin{figure}
\plotone{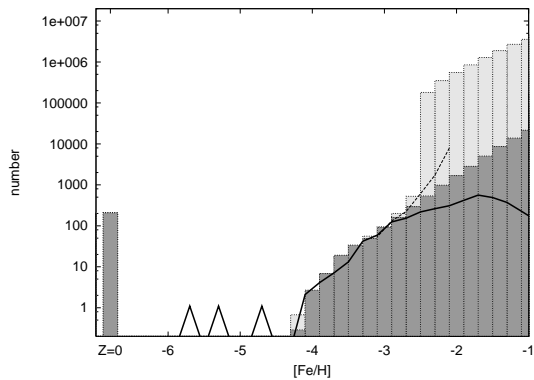}
\caption{
Basic features of metallicity distribution function (MDF) of EMP survivors. 
   Shaded histogram denotes the number of stars predicted by our fiducial model (F0) with a high-mass IMF of the medium mass $\mmd=10 \msun$ and the dispersion $\dm= 0.4$ for the same area coverage and efficiency as HES survey.
  Light gray histogram denotes the predicted number when we change the IMF into a low-mass IMF of Galactic spheroid \citep{Chabrier03} for $\feoh > -2.5$ (Fs).  
   Solid line shows the observed MDFs for EMP stars by the HES survey \citep{Beers05} and for the kinematically selected Galactic halo stars \citep{Ryan91}, matched to the HES survey for $\feoh \gtrsim -2.5$. 
   For $\feoh < -3$, plotted is MDF of stars with HDS follow-up observation, compiled by the SAGA database \citep{Suda08}.  
   Dashed line shows the ``corrected" MDF of the HES survey by \citet{Schorck08}.   
}
\label{MDF}
\end{figure}

We first discuss Model~F0; our fiducial model with the same high-mass IMF and SFE applied to the stars of all the metallicity from Pop III and EMP stars to Pop II stars of $\feoh >-2.5$ without any feedback.  
   The resultant MDF is shown (shaded histogram) and compared with observations in Figure~\ref{MDF}.  
   Here the model MDF is given under the assumption of the same area coverage and efficiency as HES survey.  
   In the hierarchical structure formation scenario, the stars born before or after the first pollution constitute two separate groups.  
   The stars, born after the first pollution, reproduce the observed MDF in the metallicity range $ - 4 \lesssim \feoh \lesssim - 2.5$.  
   In particular, the predicted number of stars agrees well with the observed number of EMP survivors.  

Our model well represents the observations in the total number of EMP survivors as well as in the shape of the MDF.  
   This is due to high mass IMF for EMP population. 
   The model with low-mass IMF predicts a much larger frequency of survivors as shown in the light gray histogram in Fig.~\ref{MDF},  and is inconsistent with the observations. 

We present the results of computations for one merger tree because results are almost the same for any merger tree.  
   This is because many low-mass mini-halos take part in the formation of EMP stars.  
   Merger trees with $10^{12}\msun$, have $\sim 2 \times 10^5$ branches with $10^6\msun$ and $\sim 4 \times 10^3$ branches with $10^8\msun$. 
   The history of a major merger in low redshift is different from tree to tree but the distribution of low mass mini-halos is similar because of the large number.  
   For computation with WMAP5 parameter, merger trees have smaller branches and predicted number of Pop.~III stars decreases with $\sim10\%$. 
   

\begin{figure}
\plotone{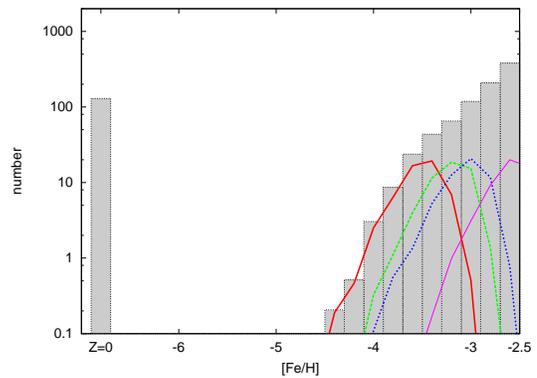}
\caption{
   Frequency distribution of metallicity of EMP survivors, descendant of SNe in each mini-halo. 
   Thick solid (red) line shows MDF of stars, born between the first pollution and the second SNe in their host mini-halo. 
   These stars are formed out of gas mixed with the ejecta of only one SN and are distributed around metallicity slightly larger than $\feoh \sim -4$. 
   Dashed, dotted and thin solid (green, blue, magenta) lines denote the number of EMP survivors, descendant of the 2nd, 3rd and 10th SNe, respectively.  
}
\label{generation}
\end{figure}

In our model, the break of MDF near $\feoh \simeq -4$ is reproduced without recourse to any ad-hoc change of IMF.  
   This is the natural consequence of hierarchical scenario as discussed in \S~\ref{cutoffS}.  
   Figure~\ref{generation} shows the number of EMP survivor descendants measured by the number of SNe until they acquire their metallicity. 
   The distribution of EMP stars, formed after the first SNe and before the next SNe in each mini-halos, is plotted in thick solid (red) line.  
   The 2nd-generation of stars are distributed around $\feoh\sim -3.6$ and form a steep slope of MDF around $\feoh\sim -4$.  
   Metallicity distributions of subsequent generations of stars, which are descendants of the 2nd, 3rd and 10th SNe, are also shown. 
   The metal abundance of EMP stars is approximately proportional to the number of progenitor SNe.
   Our model shows that EMP stars with metallicity $\feoh\sim -2.5$ have $\sim 10$ SN progenitors. 

For $\feoh > -2.5$, the fiducial model has a similar slope of MDF to that of EMP stars since we assume the same IMF as for EMP population. 
   This seems consistent with the observations for $\feoh \lesssim -2$. 
   \citet{Ryan91} report that the MDF smoothly extends from the peak near $\feoh \simeq -1.8$ into EMP stars for the kinematically selected halo stars.  
  In contrast, the switchover to a low-mass IMF, characteristic to Pop.~II stars, causes a great increase of low-mass survivors to be incompatible with the observations, as seen from the Fig.~\ref{MDF}.  
    The ``corrected'' sample of HES survey \citep{Schorck08} has a steeper increase of MDF for $\feoh>-3$ than the sample of \citet{Ryan91}, and yet, their MDF is also by far smaller than the result with low-mass IMF and still closer to the fiducial model.   
   This indicates that no significant change in the IMF is needed within this model as far as the field stars in the Galactic halo are concerned. 

\subsection{Pop.~III  Stars}

The fiducial model predicts a large number ($\sim 300$) of Pop.~III stars. 
   Since there are no other observational counterparts in this metallicity range, it is natural to assign them to HMP stars.  
   However only three HMP stars have been discovered to date and by far fewer even if we take into account the fraction of stars with follow-up observations.  

\begin{figure}
\plotone{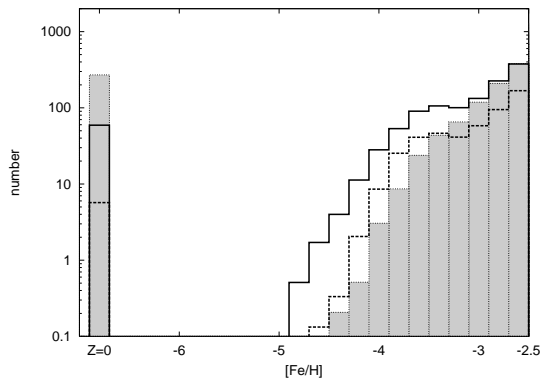}
\caption{
MDF resulting from the models under the different assumptions on the star formation for Pop.~III stars. 
   Solid and dashed lines denote results of Model~F-10 with the negative feedback by the radiation of the first stars and of Model with IMF of $\mmd=100 \msun$ for the stars with Z=0, respectively. 
}
\label{pop3}
\end{figure}

In Figure~\ref{pop3}, we show the model including local radiation feedback by solid line.    
   Here we assume that the first massive stars of $M > 10 \msun$, once born, prohibit star formation in their host halos until their explosion.
   We find that the predicted number of Pop.~III survivors decreases by a factor of $\sim 5$.  
   Accordingly, we may expect $\sim 60$ Pop.~III survivors in the HES survey area, which is still larger up to by a factor of $\sim$15 than the number of currently estimated HMP stars.  

A hump around $\feoh \sim - 3.5$ is due to the star formation after the first pollution and before the second SNe in the host halos, which may be an artifact of our assumption to take into account the radiative feedback only for the first stars.   
   In this model, the first massive star is formed similar to Model~F0, but the formation of 2nd generation stars, and hence, the second SN explosion is delayed.  
   During the elongated time interval between the first and second SNe, the second generation stars are formed as many as $\sim 5$ times larger compared with Model~F0.   
   In actuality, however, radiative feedback should work for the second and subsequent generations of stars, and hence, the hump will spread to be alleviated if we include this effect.  

For models with smaller $\sigma_8$ by WMAP 5-year, the predicted number of Pop.~III stars becomes $\sim 10 \%$ smaller because of a smaller number of merger tree branches, but it is still much larger than the observed number. 

The MDF predicted by the Model~F100 with very massive Pop.~III stars is also illustrated in this figure (dotted line).  
   For Pop.~III stars, we assume the IMF with $\mmd =100\msun$, and the same set of parameters as the fiducial model without radiation feedback.   
   Since we assume $n(q)=1$, the fraction of low-mass star companions decreases in inverse proportion to the medium mass. 
   In addition, the total number of Pop.~III stars is also reduced because SFE is defined as the total mass of stars formed in unit time in unit mass. 
   As a result, the predicted number of Pop.~III survivors is reduced by factor of $\sim 1/100$ to be comparable to the number of observed HMP stars. 
   However, in this case, significant fraction of stars becomes PISNe and the assumption without gas blown off the mini-halos is not realistic. 
   We will discuss feedback effects with gas blown-off the mini-halos in \S ~\ref{firstS}.
   In this case, the number of stars with $-4 \lesssim \feoh \lesssim -3$ becomes larger than the fiducial model, too, 
because of a smaller number of Pop.~III stars than in the fiducial model.  
   It is also possible to explain the scarcity of HMP stars in terms of lower binary fraction or different mass ratio distribution. 
   The formation of the first low-mass stars and the origin of HMP stars will be discussed again in the following sections.

\subsection{Dependence on the Star Formation Efficiency}\label{sfeS}

\begin{figure}
\plotone{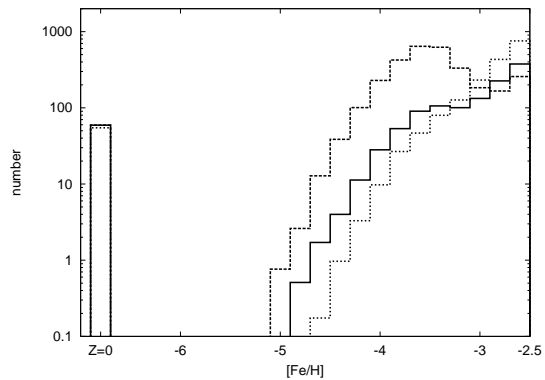}
\caption{The dependence of the MDF on star formation efficiency (SFE) for the model with radiative feedback. 
   Dashed, solid, and dotted lines show the results for Model~F-9, F-10 and F-11 with $\sfe = 10^{-9}$, $10^{-10}$ and $10^{-11} {\rm yr^{-1}}$, respectively. 
   There is a remarkable hump around $\feoh\sim -3.5$ in the model with high SFE because the number of the second generation stars formed between the first pollution and the second SNe in their host halo increases with the star formation rate ($\propto \sfe$).  
   Difference in MDF reduces for higher metallicity.  
}
\label{SFE}
\end{figure}

SFE is an important parameter for star formation and chemical evolution and yet to be properly revealed in the very high redshift and/or in very metal poor environment.  
   It could be different from SFE in the nearby universe and is free parameter in our computations.  
   Figure~\ref{SFE} compares the results of Models~F-9, F-10 and F-11 with different SFE, $c_\star =10^{-9}$, $10^{-10}$ and $10^{-11}$, respectively. 
   For the models of two larger SFE, a bump on the MDF develops around $\feoh\sim -3.5$ in proportion to SFE, as stated above. 
   For the model of the lowest SFE, F-11, on the other hand, no bump is discernible since the interval between the successive star formation grows as large as the lifetime of massive stars, $\tau \simeq 10^7$ yr.  
   Apart from the bump, the MDFs are almost independent of SFE not only for Pop.~III stars but also for the stars of subsequent generations.  
   The reason for this is as follows;  
   if we increase SFE, the number of stars increases at every timestep in proportion to SFE.  
   However, at the same time, the interval of SNe, and hence, the timescale of chemical enrichment become smaller.  
   This will reduce the number of stars at given metallicity in inverse proportion to SFE. 

In these computations, we assume that radiative feedback works only for Pop.~III stars.  
   After the first pollution, however, massive stars also dissociate ${\rm H}_2$ molecule and ionize gas in the host halo to decrease the SFE.  
   If we assume that one massive star prohibits star formation in H II region with mass $M_{\rm H II}$, the star formation dies down when the prohibited region covers the whole gas mass of the host halo, i.e.,  
\begin{equation}
\frac{M_{\rm gas} \sfe f_{\rm MS} \tau_{\rm MS}} {\overline{M}}  M_{\rm H II} > M_{\rm gas}, 
\end{equation} 
   where $\tau_{\rm MS}$ is the lifetime of H II region, $f_{\rm MS}$ is the number fraction of massive stars, and $\overline{M}$ is the averaged mass of stars.  
   Thus the effective star formation efficiency is thought to be regulated to   
\begin{eqnarray}
\sfe & \lesssim & \frac{\overline{M} }{ f_{\rm MS} M_{\rm H II} \tau_{\rm MS}} \nonumber \\
& \sim & 3\times 10^{-11} \left( \frac{M_{\rm H II}}{10^5\msun} \right)^{-1} \left( \frac{{\tau}_{\rm MS}} {10^7 {\rm yr}} \right)^{-1} {\rm yr}^{-1}, 
\end{eqnarray}
under the high mass IMF with $\mmd=10\msun$.  
   This is slightly smaller than what we assumed in our fiducial model and seems to be consistent with observed MDF without hump around $\feoh \sim -3.5$.  
   It is true, on the other hand, that some positive feedback from massive stars is thought to enhance the star formation.  
   In the relic site of H II region, the formation of ${\rm H}_2$ molecule is argued to increase the formation rate \citep{Ricotti02}.    
   The possibility is explored that SN blast wave shell may trigger the subsequent star formation, particularly of low-mass stars \citep{Uehara00,Machida05}. 
    These possible effects of positive feedback have to be considered in further discussion of star formation efficiency in the very beginning of chemical evolution.  
   In any case, the scarcity of HMP stars and the absence of a hump near $\feoh \simeq -3.5$ in the observed MDF indicate smaller star formation efficiency than $c_\star =10^{-10} {\rm yr}^{-1}$ assumed in our fiducial model for the first and subsequent star formation.   
   
In the following, we show results of models with $\sfe=10^{-11}$. 

\subsection{Chemical Evolution and Formation History of Galactic Stellar Halo}\label{history}

\begin{figure}
\plotone{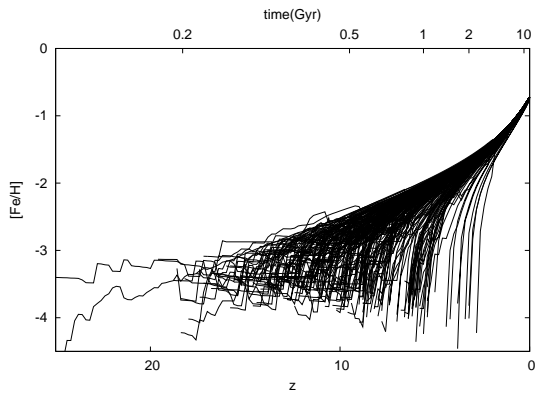}
\caption{
Chemical evolution along the merger tree for Model~F-11.  
   Most of the EMP stars are formed in high redshift of $z \gtrsim 10$.  
   There is a large diversity of metallicity of halos in the high redshift and EMP stars are formed in many mini-halos that collapse in various redshift. 
}
\label{CE}
\end{figure}

\begin{figure}
\plotone{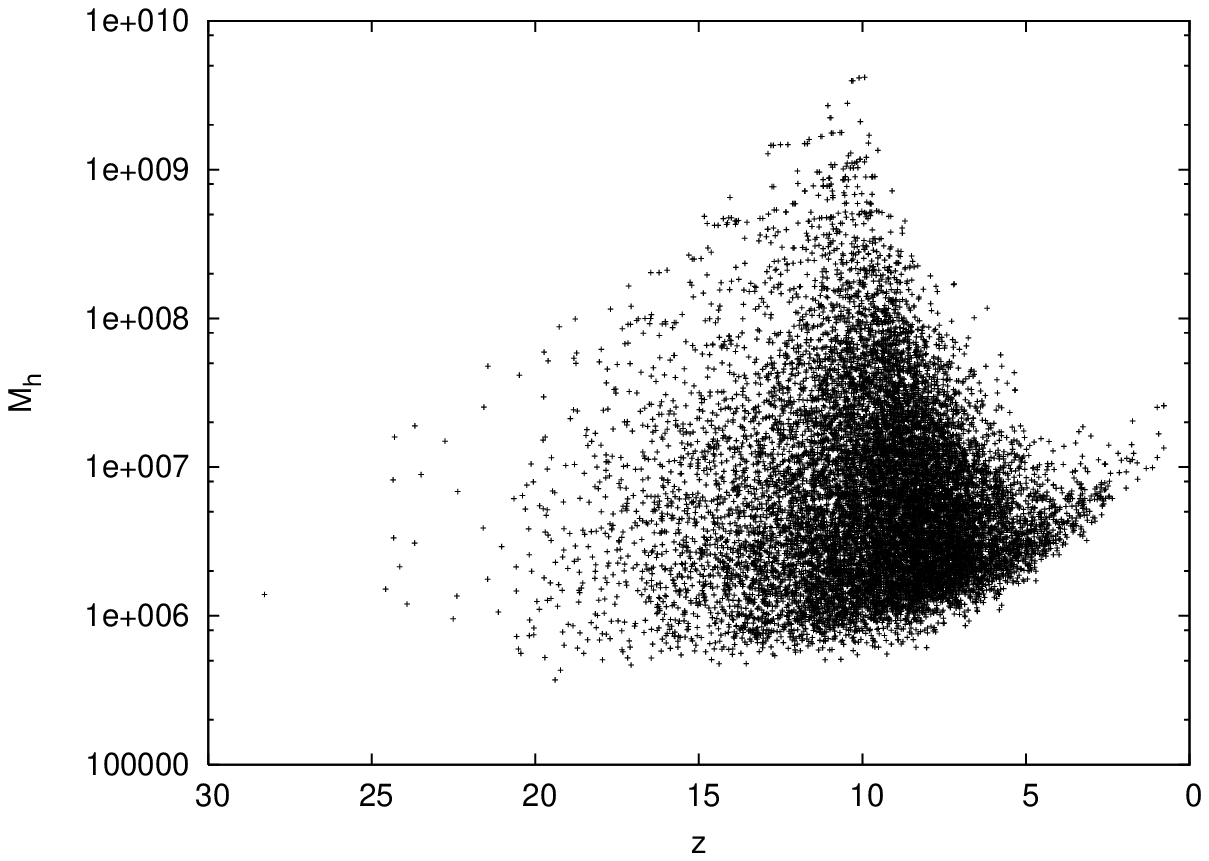}
\plotone{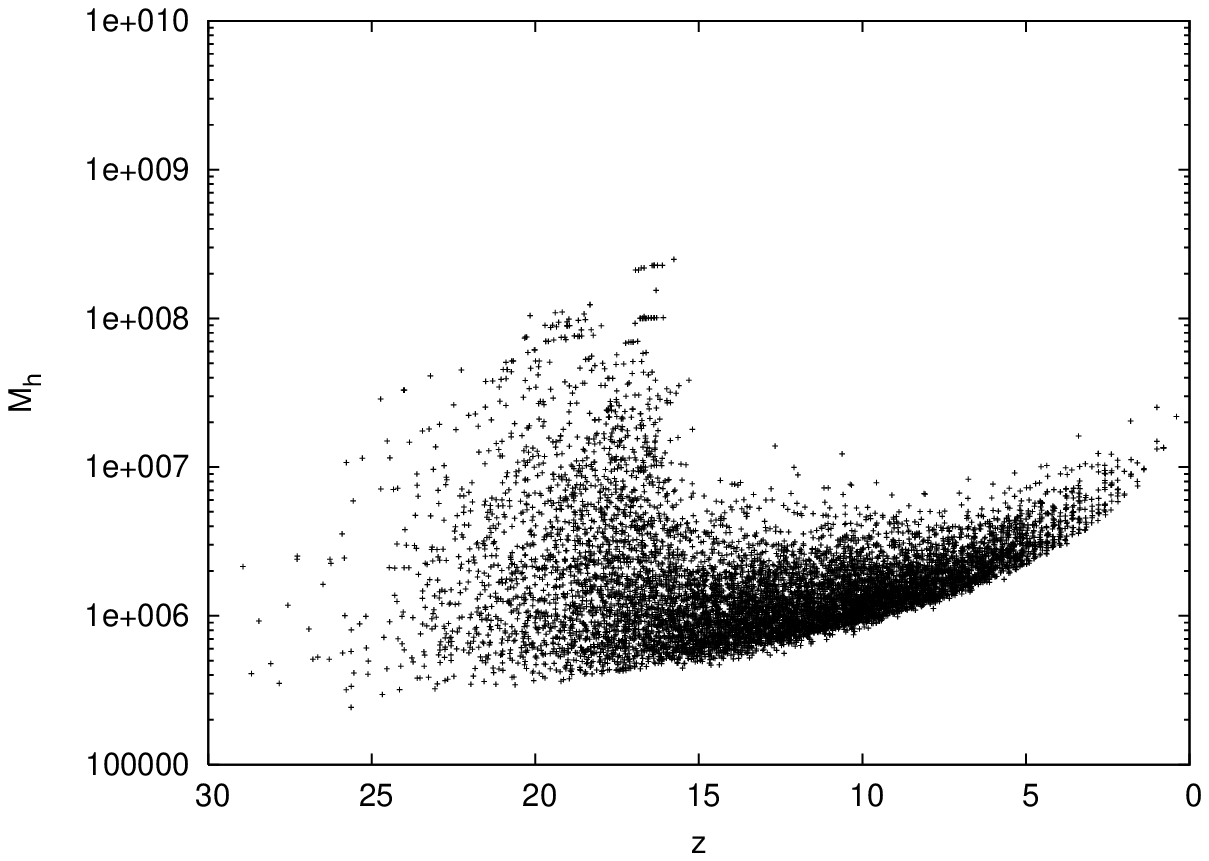}
\caption{
Distribution of the formation redshift of EMP stars with $\feoh<-2.5$ and the mass of their host halos at their birth for Models F-11 (top panel) and F-10 (bottom panel), respectively. 
}
\label{zM}
\end{figure}

\begin{figure}
\plotone{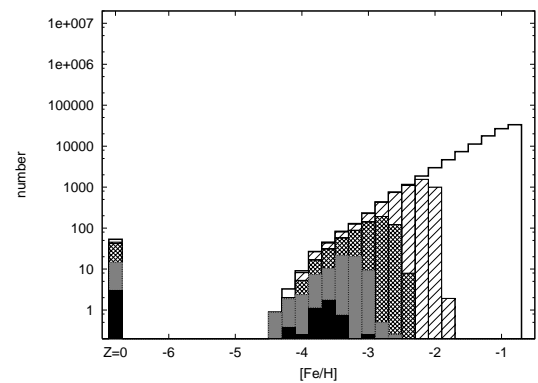}
\caption{
Evolution of the MDF with the redshift.  
   Black, gray, cross-hatched, and hatched histograms denotes the MDFs of EMP and Pop.~III survivors formed before $z = 20,$ 15, 10, and 5 , respectively.  
   Thin solid line denotes MDF at z=0. 
   The number of stars in the MDF increases with time, while its shape is almost unchanged for $\feoh<-2.5$ . 
}
\label{red}
\end{figure}

Figure~\ref{CE} illustrates the evolution of metallicity of sampled mini-halos against the redshift and age of the universe for Model~F-11.  
   Most mini-halos are formed in $20 > z >5$ and they evolve to $\feoh > -2.5$ in timescale for $\sim10$ SN explosions, as shown in Fig.~\ref{generation}.  
   When we neglect accretion of inter-halo matter, the timescale, $t_{\rm enri}$, to enrich the mini-halos with metal up to $\feoh$ is given by 
\begin{eqnarray}
t_{\rm enri} &=& \frac{10^{\feoh} \Zsun \overline{M}}{\sfe M_{\rm Fe} f_{\rm SN}} \nonumber \\
&\sim & 68 \times 10^{\feoh} {\rm Gyr}, 
\label{timescale}
\end{eqnarray}
   for Model~F-11, where 
$f_{\rm SN}$ is the number fraction of massive stars to be SNe, and $\overline{M }$ is the averaged stellar mass.  
   When this timescale is shorter than the collapse timescale, $ t_{\rm coll}$, of the mini-halos that enable the star formation, the chemical evolution progresses separately in each mini-halo.   
   For Model F-10 with higher SFE, the timescale is short enough for the independent evolution and for Model with F-11, some EMP stars are formed in the merged halo. 

Figure~\ref{zM} shows the distribution of EMP and Pop.~III stars on the diagram of the formation redshift and mass of host halo at their birth for Models~F-11 and F~-10.  
   Most EMP stars are formed in mini-halos with $10^6 - 10^8 \msun$ at $z\sim 5-15$. 
   As shown in Fig.~\ref{CE}, for Model~F-11, metallicity $z=0$ are 0.7 dex lower than solar metallicity, indicating that SFR is higher at least in the Galactic disk at lower redshift. 
   
We assume that all halos with $T_{\rm vir}>10^3$ host stars but the re-ionization is thought to prohibit the accretion of gas onto halos with small mass of $T_{\rm vir}<10^4$. 
   For models with higher SFE, the results about EMP stars are quite similar because most EMP stars are formed at $z > 10$ and before the reionization. 
   For Model F-11, a significant fraction of EMP stars are formed in lower redshift but the resultant MDF is quite similar even under the effect of reionization. 
   

Figure~\ref{red} shows the distributions of Pop.~III and EMP survivors that have been formed by certain period or redshift for Model~F-11.  
   As the redshift decreases from $z = 20$ to 5, the distribution of stars, formed by the given redshift, increases greatly in number and extends into higher metallicity. 
   Pop.~III stars and EMP stars are mainly formed at high redshifts of $z \gtrsim 10$ and the shape of MDF remains little unchanged afterwards.  
   In the mini-halos, stars have the metallicity of $\feoh \gtrsim -4$ apart from Pop.~III stars.  
   In the mini-halos that evolved to larger mass by accretion and/or by merger, metals ejected by SN are diluted in larger mass, and hence, the 2nd generation of stars tend to have smaller metallicity. 
   
Consequently, the metallicity of EMP stars weakly depends on the formation epoch for the hierarchical scenario. 
   There is a weak age-metallicity trend for EMP stars.
   The trend appears only when $t_{\rm enri} (\feoh) > t_{\rm coll}$.    
   As seen in eq.~(\ref{timescale}), the timescale, $t_{\rm enri}$, of the chemical enrichment becomes shorter for higher SFE, and the age-metallicity trend among EMP stars is not discernible for models with higher SFE.   

\section{Result 2: Accretion of Interstellar Matter }\label{accS}


In our models, a group of stars are formed with $Z=0$.   
   In this section, we show the results with Models F-11, FA, FB and FD, and discuss the surface pollution of these stars by the accretion of ISM.  
   In the previous section, we calculate the formation history of metal-poor stars and the chemical enrichment history of mini-halos with a merger tree.  
   Here we trace the changes of surface iron abundances through the accretion of ISM for all the individual Pop.~III and EMP survivors formed in the model calculation by taking into account the structural and chemical evolution of halos.  
   We investigate the dependence on the velocity of stars relative to the ISM, the distribution of gas in the mini-halos, and the accretion process in binary systems.  
   Based on the results, we may estimate the observed abundances of Pop.~III survivors and discuss the lowest surface metal abundances of halo stars.   

For HMP stars, the most metal-poor stars, known to date, the peculiar abundance patterns such as very large carbon enhancement have been discussed.  
   In this paper, we focus on the iron abundances, however, since the observed abundance patterns of the light elements from lithium, carbon through aluminum are investigated in detail and shown to be affected by binary mass transfer \citep[][see \S~\ref{HMPabundance} below]{Suda04,Nishimura09,Suda09b}. 
   The iron abundance is a good indicator for ISM accretion because it is not affected by the binary mass transfer and the evolution of low-mass Pop.~III survivor themselves.

\subsection{Surface Pollution History}\label{traceS}
\begin{figure*}
\epsscale{1.0}
\begin{tabular}{cc} 
\begin{minipage}{0.5\hsize}
\plotone{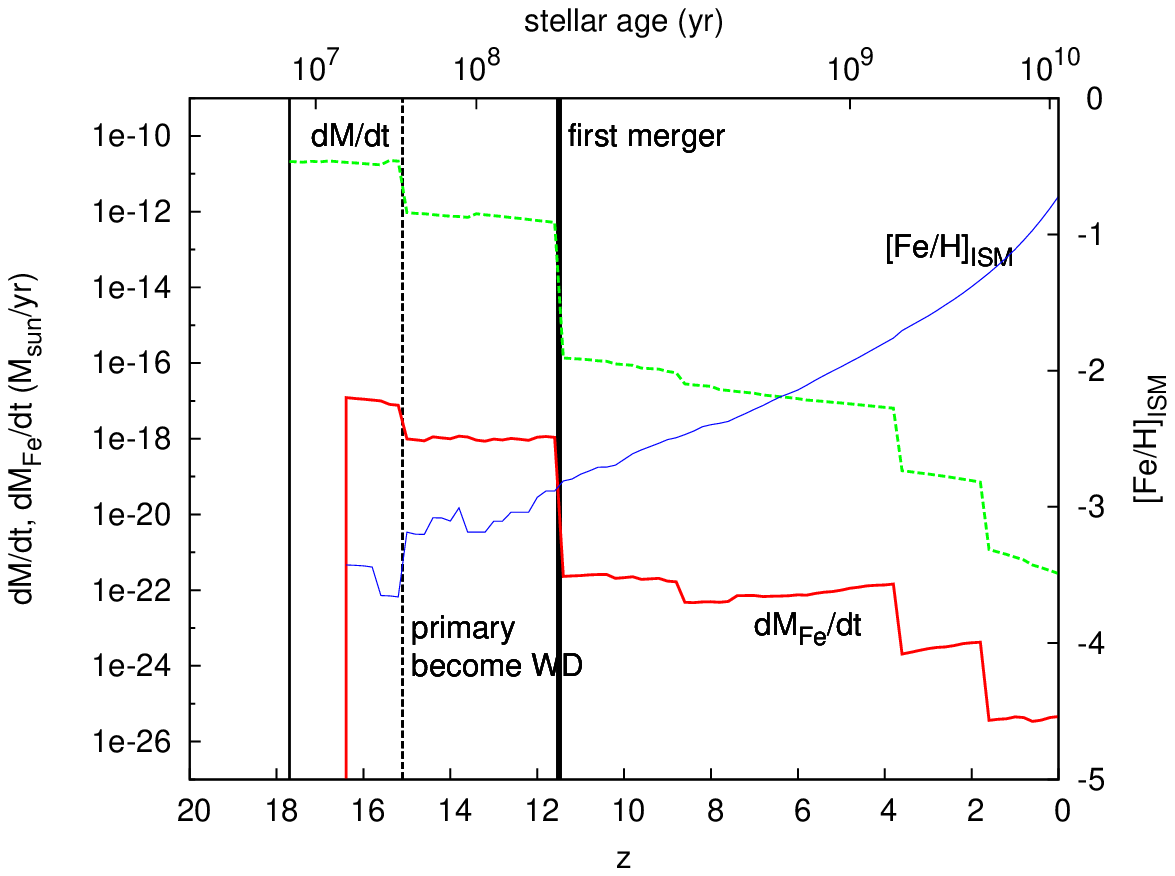}
\end{minipage} 
\begin{minipage}{0.5\hsize}
\plotone{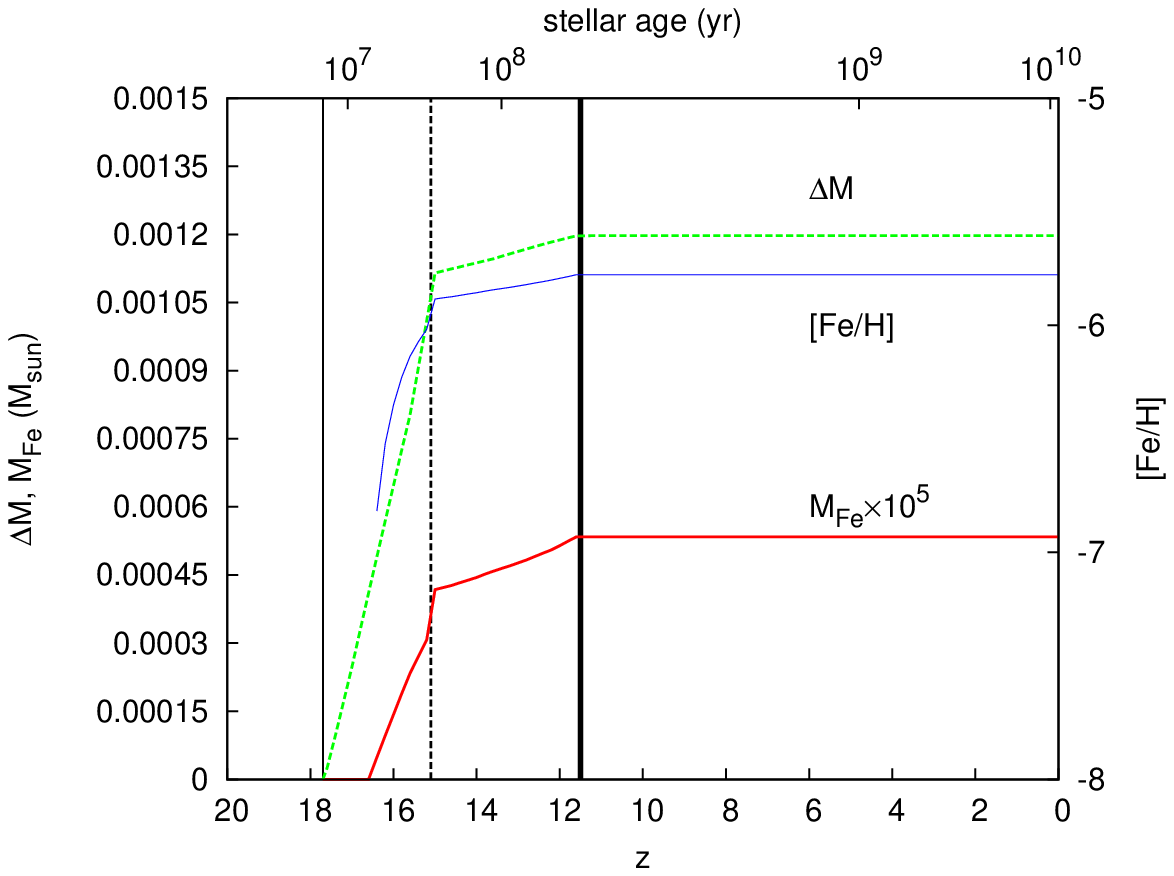}
\end{minipage} \\
\begin{minipage}{0.5\hsize}
\plotone{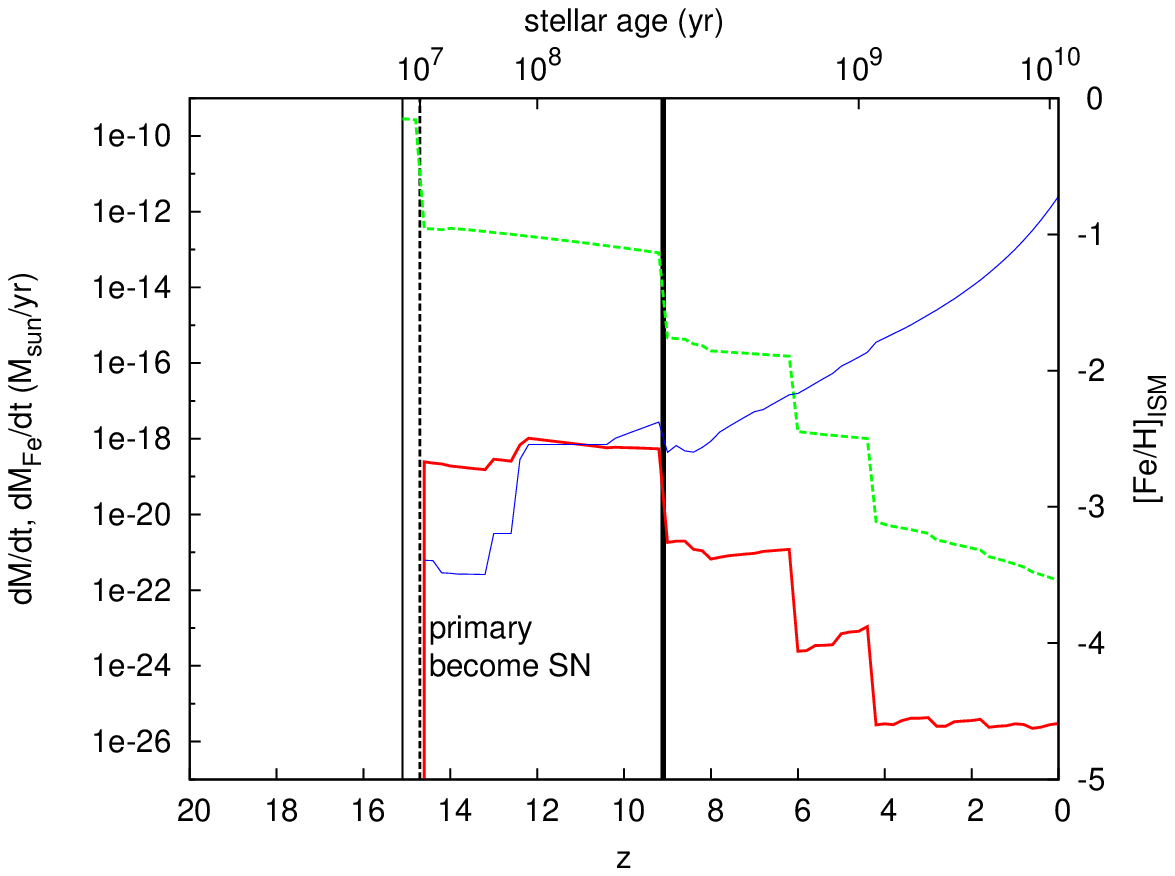}
\end{minipage} 
\begin{minipage}{0.5\hsize}
\plotone{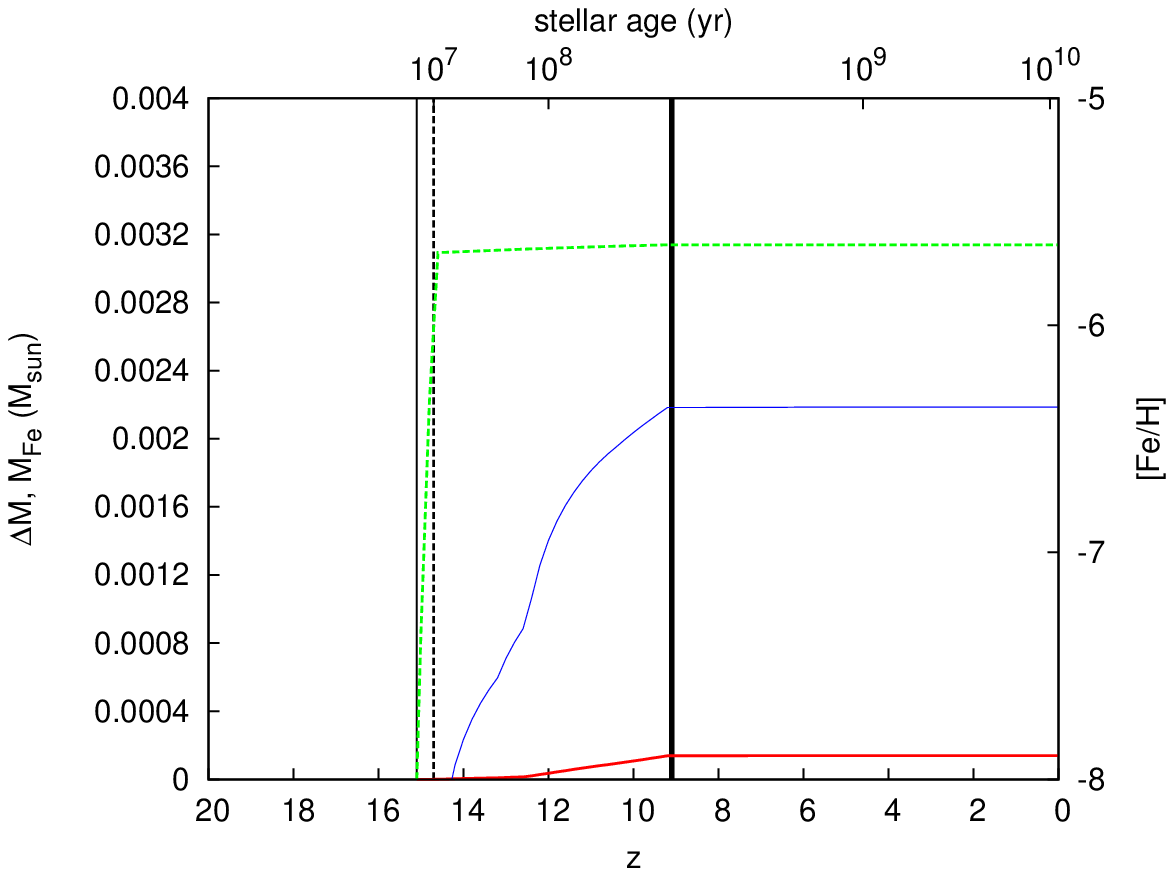}
\end{minipage} \\
\begin{minipage}{0.5\hsize}
\plotone{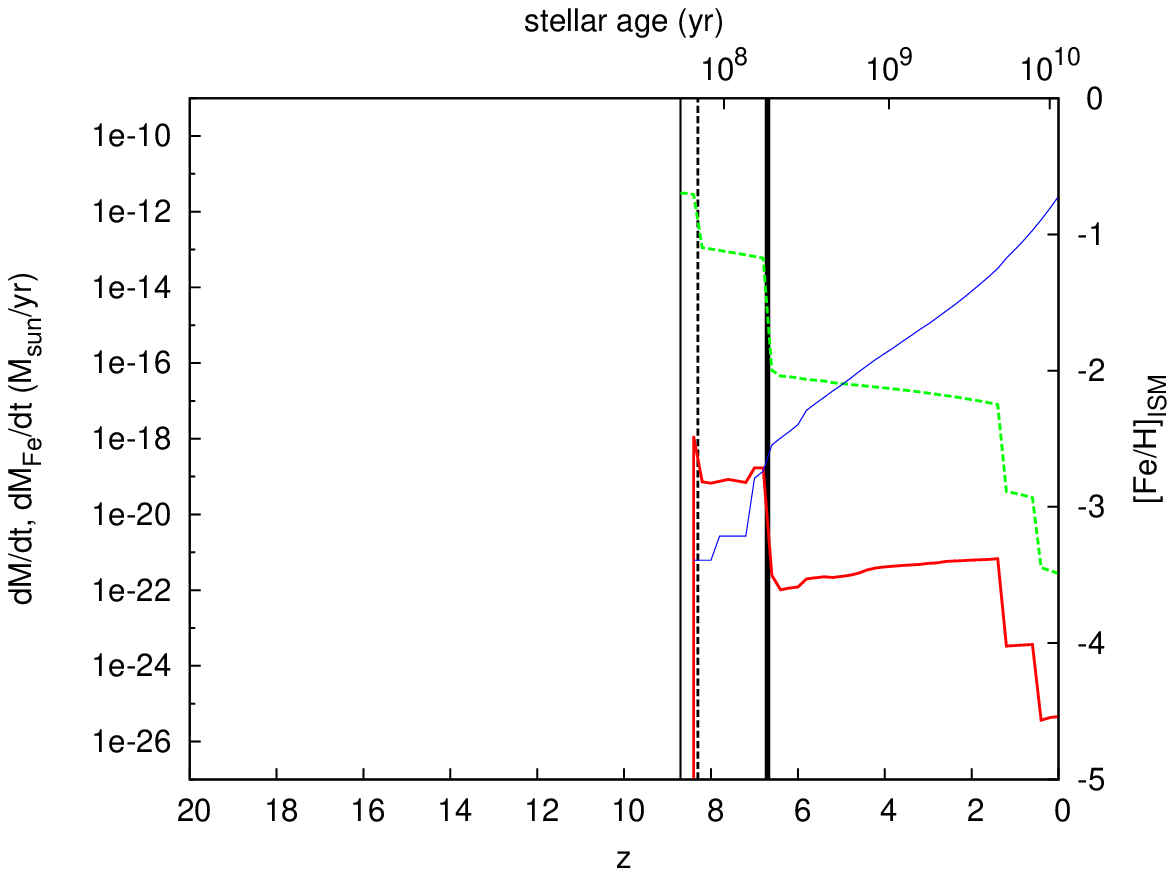}
\end{minipage} 
\begin{minipage}{0.5\hsize}
\plotone{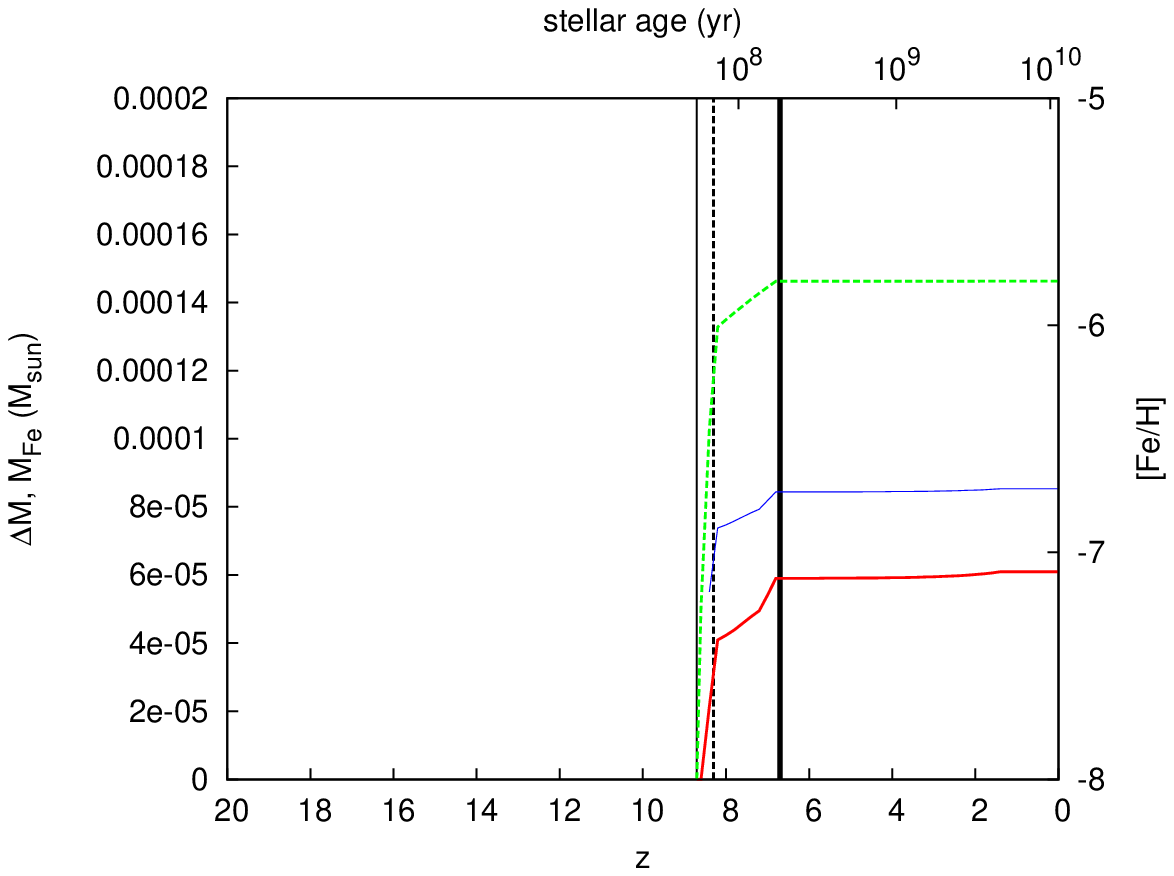}
\end{minipage} 
\end{tabular}
\caption{Surface metal pollution history of three sample Pop.~III stars of mass $0.8 \msun$ in the Model~F-11.  
   Left panels show the accretion rate of iron (thick solid red lines) with the accretion rate of gas (dashed green lines) and the iron abundance of ISM gas (thin solid blue lines) as a function of the redshift (bottom axis) and the age of the star (top axis), respectively.
   Right panels show the surface iron abundances of polluted Pop.~III giants (thin solid blue lines) with the cumulative masses of accreted iron (thick solid red lines) and gas (dashed green lines).  
   Vertical lines mark the time for the formation of the stars (thin solid lines), the merger of the birth halos of Pop.~III stars with larger halos (thick solid lines) and the end of the nuclear burning stage of their primaries (dashed lines), respectively.   
   Top, middle, and bottom panels represent the results for Pop.~III survivors, formed with primaries of mass $5.8, 29.7$, and $6.0 \msun$ at the redshift $z =17.7, 15.1, 8.7$ and the time $t=0.20, 0.30, 0.44 {\rm Gyr}$, respectively.  }
\label{trace}
\end{figure*}

We illustrate the detailed history of ISM accretion and surface pollution for the Case C model in Figure~\ref{trace}, where the time variations in the accretion rates of ISM gas and iron and variations in the surface abundances with the chemical enrichment of their host halos for three sample Pop.~III survivors in the binaries.  
   In the birth halos in which Pop.~III stars are formed, the ISM accretion rate is as large as $\dot M\sim10^{-10 \mh -12}\msun/ {\rm yr}$.  
   When the birth halos merge to larger ones, the accretion rate decreases immediately by a factor of $\sim 1/100$ or less.  
   This is because we assume that gas and stars are concentrated around the center in their birth halos while the stars come to move with the virial velocity after the first merger, which reduces the gas accretion rate by $\sim 1/56$, as seen in eq.(\ref{eq:caseB}).  
   Additionally, the mass of the host halo increases by a factor of 2 or more, which further decreases the accretion rate, as seen in eq.~(\ref{eq:mdot}).  

\begin{figure}
\plotone{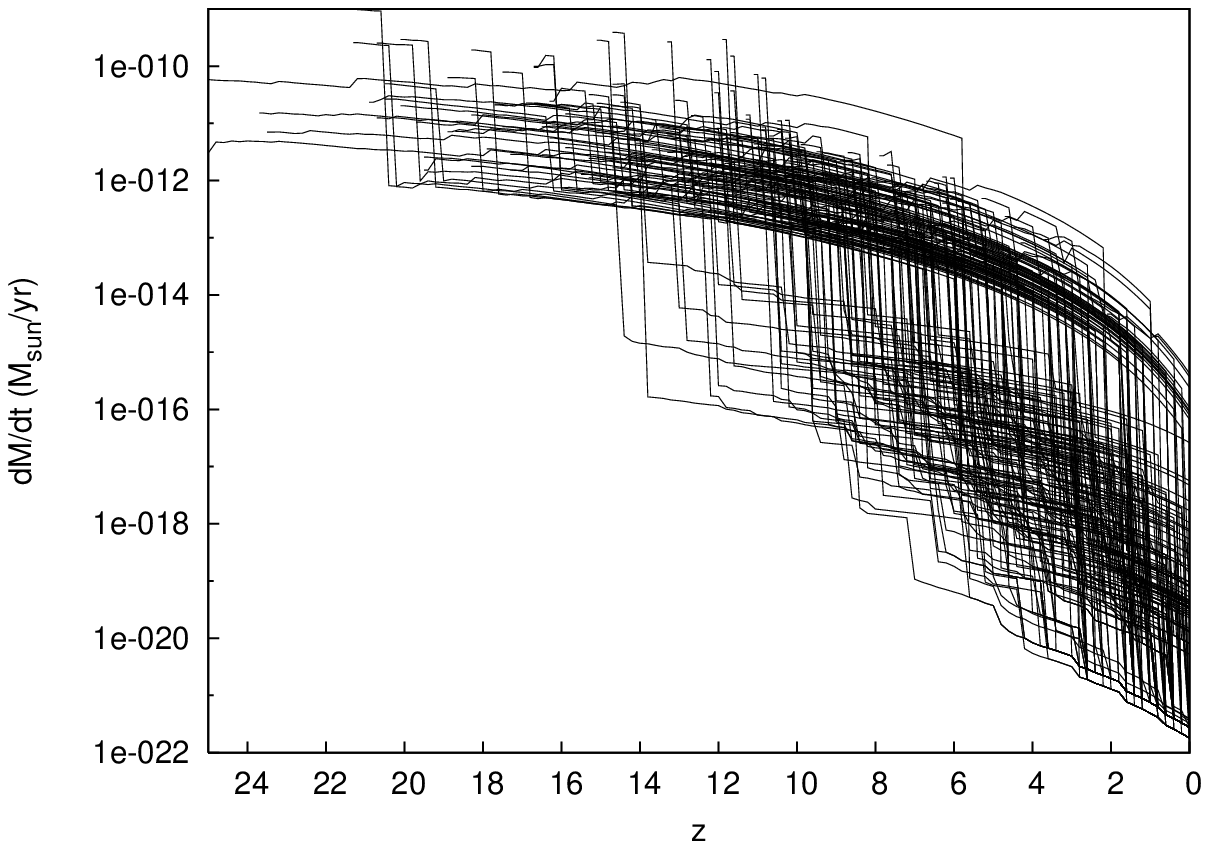}
\caption{
Variation of ISM accretion rate of $\sim 200$ sampled Pop.~III stars against the redshift, the same as dashed lines in left panels of Fig.~\ref{trace}.   
}
\label{trace2}
\end{figure}

Figure~\ref{trace2} plots the ISM accretion rate against the redshift of their formation for 200 sample Pop.~III survivors.   
   We see that the ISM accretion rate is high for mini-halos and it continues until the host halos undergo a merger with larger halos. 
   The accretion rate before the first merger is a decrease function of formation time of halos because of decrease in the $\rho_{\rm vir}$ and increase in the mass of host halos.  
   For a given redshift, it displays a spread by a factor of $\sim 10$ or more, which is caused mainly by the different masses of stars or binary systems, since the Bondi accretion rate depends on the mass (see below).   

The iron accretion rate increases from zero to $10^{-16 \mathchar`- -19} \msun \hbox{ yr}^{-1}$ after the first pollution in the birth halos. 
   As the birth halos grow by merger and the ISM accretion rate steps down, the iron accretion rate also decreases despite the chemical enrichment of the ISM with metals.  
   It decreases typically down to $\sim 10^{-25}\msun \hbox{ yr}^{-1}$ today.   
   This rate can be changed by formation of the galactic disk but the accretion rate today for halo stars with high velocity is still much lower than in mini-halos.  
   The accretion of ISM and metals is by far more effective before the first merger in the earlier universe than in larger halos in later time. 
   For Pop.~III survivors, therefore, the accretion in their birth halos is the predominant source of their surface iron abundance. 

Right panels of Fig.~\ref{trace} show the cumulative masses of accreted ISM and iron and the surface iron abundance of giant Pop.~III survivors.  
   Typically, Pop.~III stars stay in their birth clouds in $\sim 10^8$ yr before the first merger.  
   During this period, the metallicity of ISM increases from $Z=0$ to $\feoh \sim -2$ or more. 
   Pop.~III survivors accrete ISM to the amount of $10^{-2 \mh -4} \msun$ and iron to the amount of $10^{-8 \mh -10} \msun$, and which result in the surface metallicity of $\feoh \sim -4 \mh -6$. 
    After the first merger, the accretion rate becomes much lower and the surface metallicity little increases.  

The resultant surface abundance strongly depends on the merger history of host halos. 
   Between two stars in top and bottom panels in Fig.~\ref{trace}, the final surface iron abundance differs by up to about 1 dex, though the binary parameters are similar.   
   This large difference originates in the dependence of the accretion history on the formation epoch and mass of birth halos and also on the time span to the first merger. 

The accretion and pollution history also depends on the mass of erstwhile primary stars.  
   Since the Bondi accretion rate is in proportion to the mass of systems squared, the enhancement of accretion rate in the binary is large for low-mass survivors, as given by 
\begin{equation}
  \dot M \propto \frac{1}{2} \left(\frac{M_1+0.8}{0.8} \right)^2 ,  
\end{equation}
   for the low-mass member of $0.8 \msun$ stars with the primary of mass $M_1$.   
   In the top and bottom panels, we see that the accretion rate decreases about by about an order of magnitude when their primary stars become white dwarfs and reduce their mass (vertical dashed lines).   
   The iron accretion rate also decreases, though it is partly compensated by the increase in the iron abundance in ISM.  
   Since the enhancement of accretion rate ends either when the primary stars lose their masses or when the host halos undergo the first merger, there is the optimal mass of intermediate-mass primary stars that can bring about the maximum amount of accreted iron:   
   For the primary stars that have smaller masses than the optimal mass, whose lifetimes are longer than the time span to the first merger of $\sim 10^8$ yr, the amount of accreted iron increases with the mass of primaries as given above.  
   For those primary stars that have larger masses and become white dwarfs before the first merger, the amount of accreted iron tends to be smaller because of stronger mass dependence of their lifetime.  
   The difference of primary mass considerably accounts for the difference in the resultant surface iron abundances of these two stars.   

For SN binaries, the accretion rate starts with very high rates, as seen in middle panels of Fig.~\ref{trace}. 
   This is also the case for the accretion rates exceeding $\dot{M} = 10^{-10} \msun \hbox{ yr}^{-1}$ in Fig.~\ref{trace2}.   
   It has little to do with the accretion of iron, however.  
   Since the primary stars are likely to be the first massive stars in the halos, no SN is expected to pollute ISM of their birth halos until the explosion of their primary stars.  
   Before the SN explosion, gas is accreted onto Pop.~III stars but no metal accretes.   
   When the primary stars explode as SNe, the secondary stars have to be released from the binaries because of sudden decrease in the mass of primary stars, and then, begin to accrete metals as single stars.  

\subsection{Metallicity of polluted Pop.~III stars}
\begin{figure*}
\epsscale{1.0}
\begin{tabular}{cc} 
\begin{minipage}{0.5\hsize}
\plotone{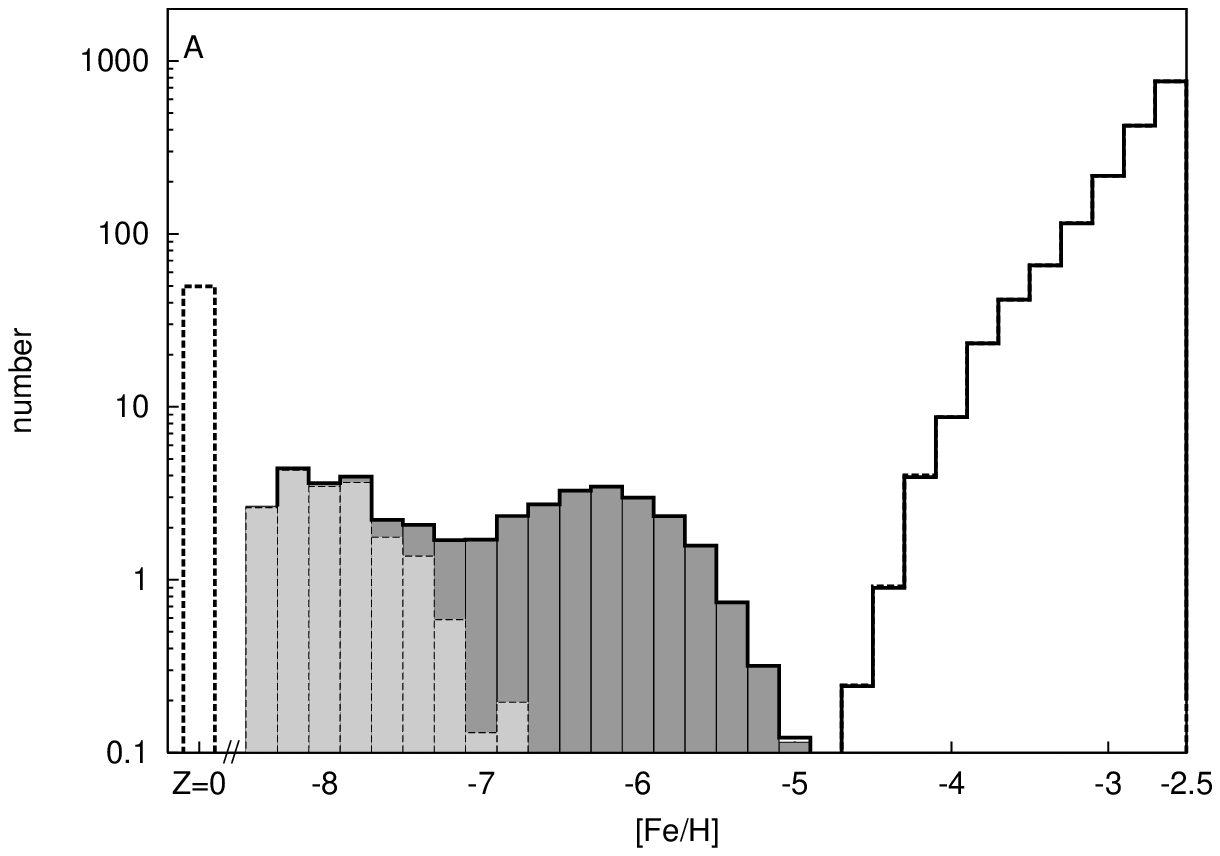}
\end{minipage} 
\begin{minipage}{0.5\hsize}
\plotone{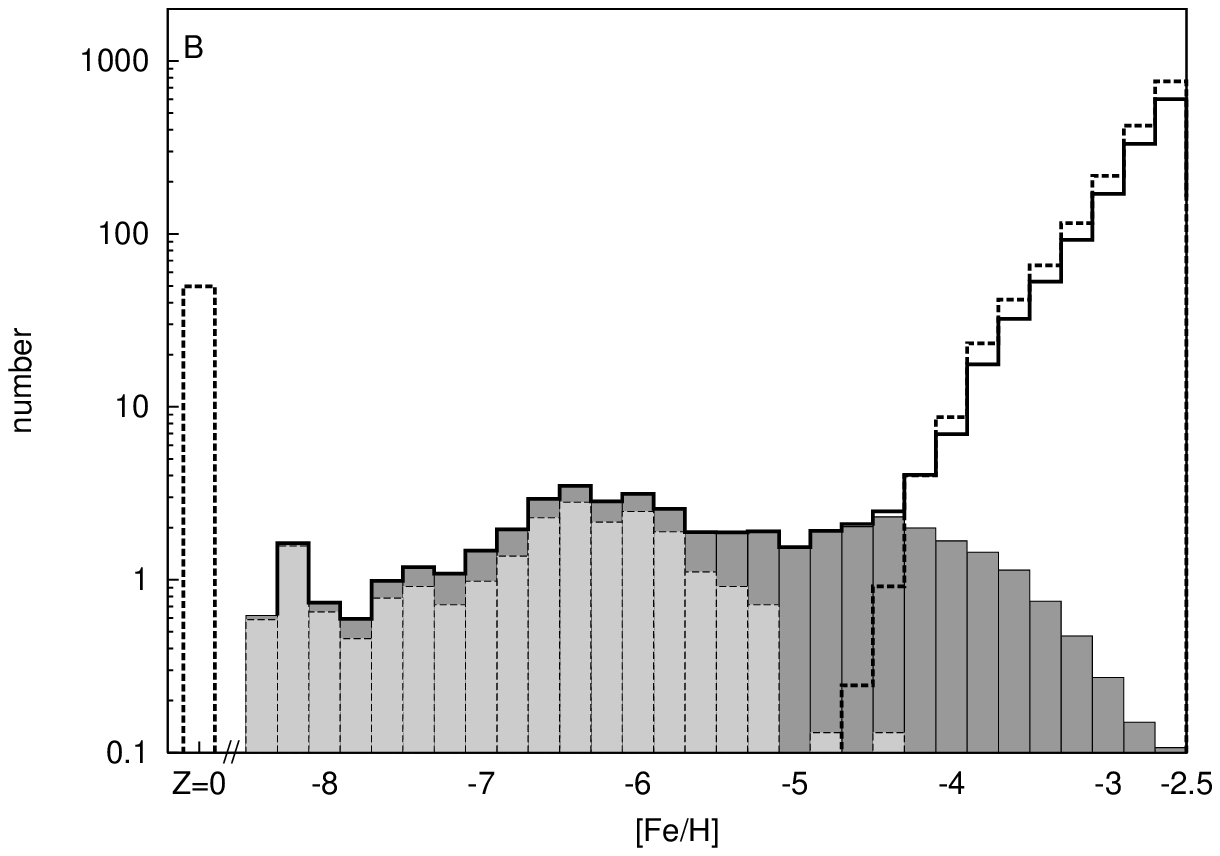}
\end{minipage} \\
\begin{minipage}{0.5\hsize}
\plotone{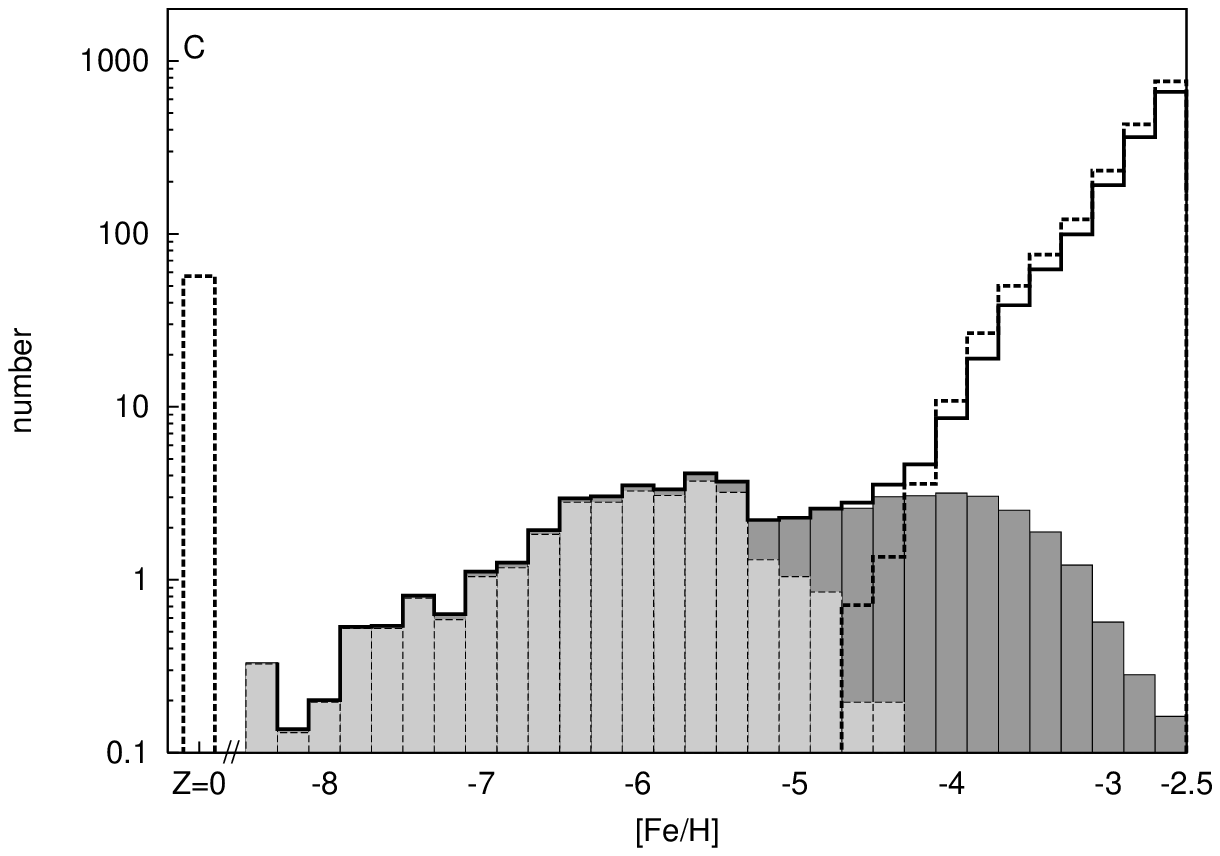}
\end{minipage} 
\begin{minipage}{0.5\hsize}
\plotone{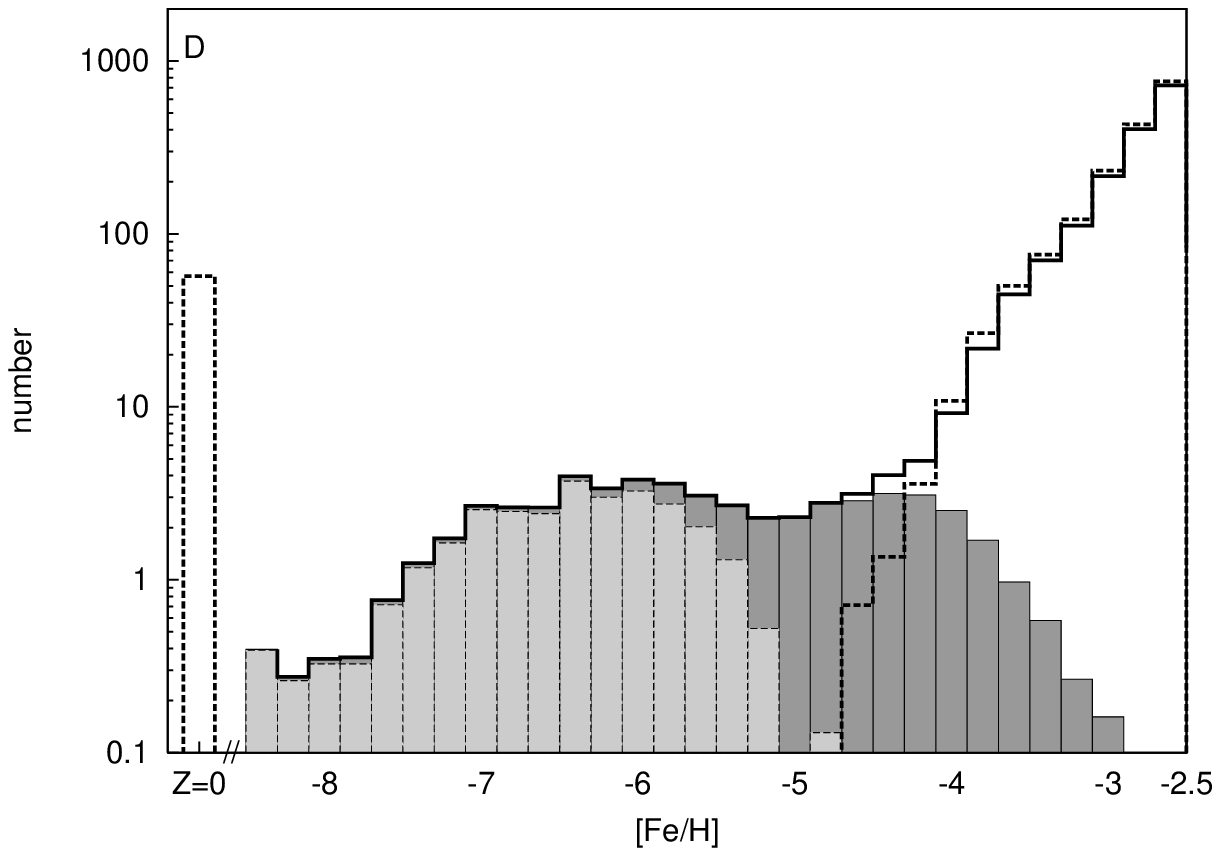}
\end{minipage} 
\end{tabular}
\caption{
   MDFs of Pop.~III and EMP stars after the surface pollution due to the ISM accretion (solid lines) in comparison with those without the ISM accretion (dashed lines).  
   Dark and light gray histograms denote the distributions of polluted Pop.~III stars for dwarfs and giants, respectively.  
   Four panels compare the models computed under the different assumptions on the dynamics of merger and accretion processes. 
   {Case A}:  Stars are assumed to move with the virial velocity in the ISM of the virialized gas density of halos (top-left panel). 
   { Case B}:  Stars are assumed to move in the cooled and centrally concentrated gas clouds at the velocity equal to the sound velocity of gas in their mother halos (top-right panel). 
   { Case C}:  Delay of merging of stars and gas in dynamical friction timescale is taken into account in Case B model (bottom-left panel):  
   { Case D}:  The members of binary systems are assumed to accrete mass of ISM independently in Case C model (bottom-right panel). 
}
\label{accretion}
\end{figure*}

In this subsection, we discuss the differences of the polluted surface abundance in four models of Cases A-D.  
   Figure~\ref{accretion} shows the MDFs with the surface metal-enrichment due to the ISM accretion for these models.  

For Case A, the surface metallicity of polluted Pop.~III star is $\feoh_{\rm{pol}}\sim-6$ for main sequence stars and $\sim-8$ for giants. 
   The accretion rate is the smallest among the four models since the relative velocity between the stars and gas in their birth halos is taken to be larger than for the other models.   
   Even in this case, the accretion in the earlier universe is dominant because of smaller virial velocity of birth halos, collapsed earlier. 

For the models of Cases B-D, the ISM accretion enriches the surface of dwarf Pop.~III stars with metals up to $\feoh \sim-3$.  
   When evolving to giants, the accreted metals are diluted to $\feoh\sim -5 - -6$, comparable to the metallicity observed for HMP stars. 
   Two of three known HMP stars are giants and we may regard them as Pop.~III stars with the surface pollution.  
   For a subgiant HE1327-2326, the surface convection is shallow and likely to be occupied by the matter transferred from the primary in which the accreted iron is diluted in the AGB envelope.  

The comparison of the models between Cases A and B reveals the dependence of accretion rate on the concentration of gas and on the velocity of stars in the birth halos of Pop.~III stars.  
   As shown in eq.(\ref{eq:caseB}), the surface pollution rate differs by $\log \ ( 200{\rm K}/1000{\rm K})^{-2.5} \sim 1.7$ dex.  
   Because the accretion in the birth halos is dominant, the final metallicity also differs by $\sim 1.7$ dex between Case A and Case B.  

Time delay of merger process increases the accreted irons by some amount. 
   For Case B, the metallicity distribution has low-metallicity tail stretching significantly beyond $\feoh \simeq -8$ while most stars fall in the metallicity of $\feoh > -8$ for Case C.  
   In the models of Case B, some mini-halos merge to a larger halo in short timescale from their formation, and hence, Pop.~III stars formed in such halos of short lives accrete little ISM and remain with very low surface metallicity.  
   On the other hand, in the model of Case C, all the Pop.~III stars stay in the same circumstances and continue to accrete ISM at the same rate as in the host halos during the dynamical friction timescale to raise the lowest metallicity.  

From the comparison between the results for Cases C and D, we see that the accretion rate rather weakly depends on the fraction of ISM that goes to secondary.  
   As discussed in the previous subsection, the enhancement of iron accretion in the binary is limited only to the primary stars of intermediate and smaller masses;  
   there is a optimal mass of intermediate-mass primary stars that bring about the largest effects, while the effect decreases both for larger and smaller masses.  
   Accordingly, the enhancement of iron accretion in the binaries ranges from several to several tens, but is not so large on average.  

In summary, the accretion of ISM can pollute Pop.~III giant stars to $\feoh \sim -5$ or more if the concentration of gas and stars in the central region of mini-halos is taken into account.  
   HMP stars can be polluted Pop.~III survivors whose surface abundances of iron group elements are strongly affected by ISM accretion.  
   The surface pollution is dominated over by the accretion in the birth mini-halos in the early universe, and this indicates that we have to take into account the hierarchical structure formation history in order to estimate the effects of surface pollution.      
   Some dwarf stars with $-4 \lesssim \feoh\lesssim -3$ also can be polluted Pop.~III stars but the number of those is much smaller than EMP stars with pristine metals.  

\begin{figure}
\plotone{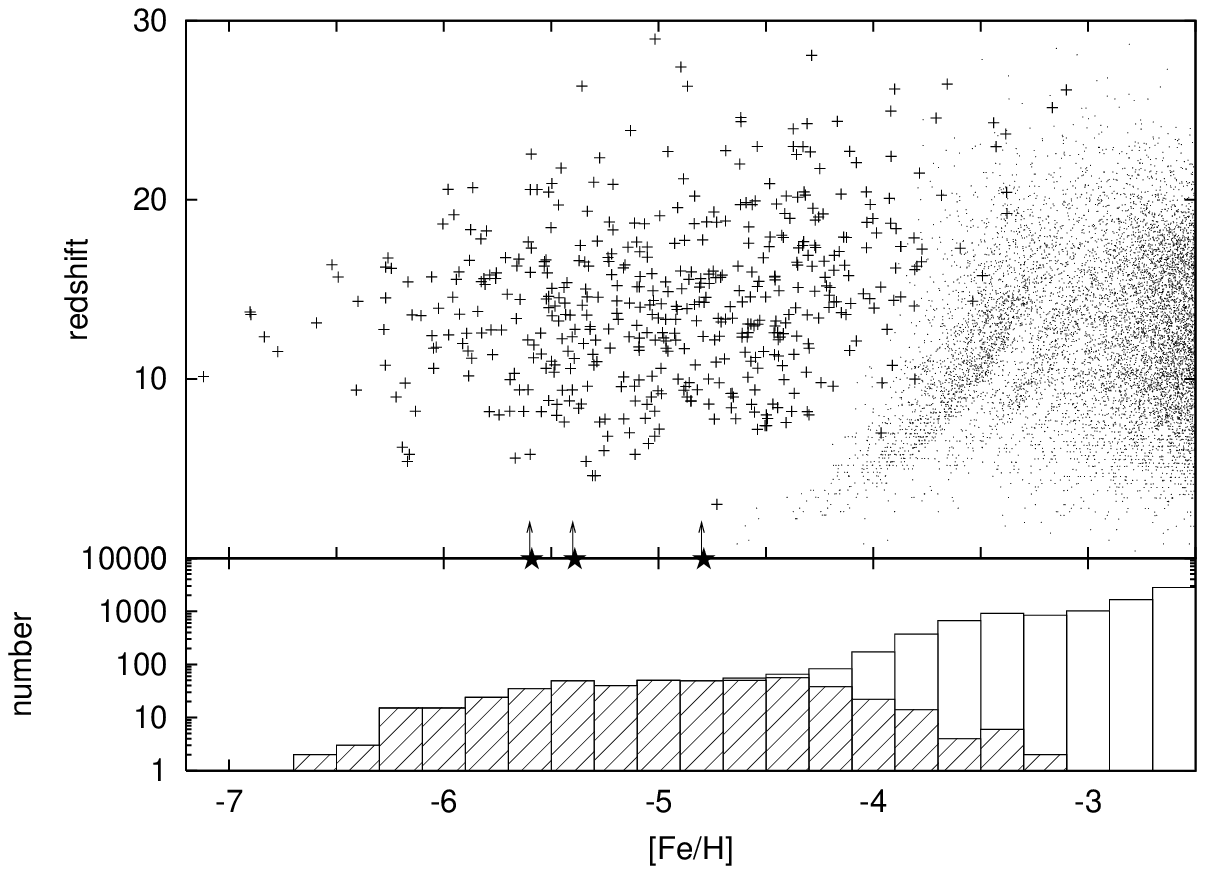}
\plotone{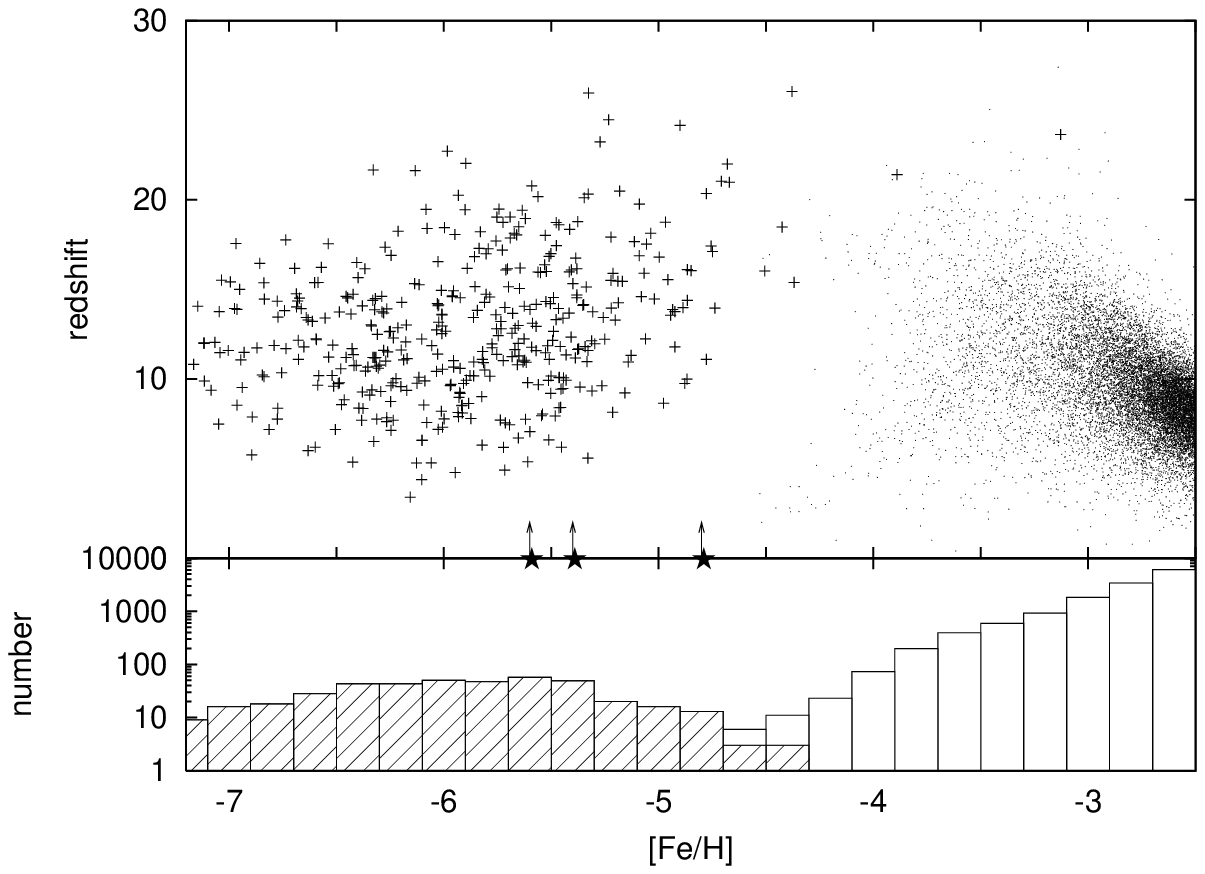}
\caption{
Distributions of the giant Pop.~III (+) and EMP ($\cdot$) survivors on the diagram of the formation redshift, $z$, and surface metallicity, $\feoh$, after the surface pollution (upper section), and their metallicity distribution (lower section).  
   Top and bottom panels show the result for the models with the star formation efficiency $\sfe =10^{-10} \hbox{ yr}^{-1}$ and $10^{-11}  \hbox{ yr}^{-1}$, respectively, and asterisks denote the iron abundances of three HMP stars.  
}
\label{zFe}
\end{figure}

Finally, we show the surface metallicity and formation redshift of polluted Pop.~III and EMP stars for the model of Case C and for the same model but for $\sfe=10^{-10} {\rm yr}^{-1}$ instead of $\sfe=10^{-11} {\rm yr}^{-1}$ in Figure~\ref{zFe}. 
   The surface pollution rate is almost independent of the formation redshift, and in proportion to SFE.  
   In addition, there is little difference in the formation redshift between Pop.~III stars and EMP stars, in particular for the model of larger SFE of $=10^{-10}{\rm yr}^{-1}$:  
   for the model of lower SFE of $=10^{-11}{\rm yr}^{-1}$, the formation of EMP stars is slightly delayed.  

\section{Result 3: Feedback Effects and Pre-pollution of IGM}\label{firstS} 

Primordial stars, formed prior to EMP stars, are thought to be very massive and to have global feedback effects on subsequent star formation.  
   Metal abundance in the most metal-deficient stars can originate in the pre-pollution of IGM by these stars.  
   In this section, we show the results of models with feedback from primordial stars and discuss the effect of pre-pollution on HMP and EMP stars.  

Table~\ref{Tresult} summarizes model results for the pre-pollution and formation of HMP stars. 
   Observationally, 3 HMP stars are discovered and several more HMP stars are expected in HES survey field if we consider the fraction of stars with HDS follow-up observations and the heavy pollution effects on main sequence stars. 
   The first column, $N_{4.5}$, denotes the expected number of stars, formed with the metallicity below $\feoh=-4.5$. 
   The second column, $\feoh_{IGM}$ denotes the final IGM metallicity by pre-pollution of blow off metals by supernova. 
   The third column denotes the fraction of iron, originated from PISN ejecta, among all the iron ejected to IGM. 
   Observational absence of nucleosynthetic signatures of PISN ejecta indicates that metal from PISNe should not be the dominant source of metal for pre-pollution. 
   
\begin{table}
\begin{center}
\caption{Model results}
\label{Tresult}

\begin{tabular}{l|rrr}
\hline
name  & $N_{4.5}$ & $\feoh_{\rm IGM}$	& $\rm Fe_{PISN}/Fe_{IGM}$ \\
\hline
F0	& 210	& /	& /	\\
F-10& 46	& /	& /	\\
F-11& 42	& /	& /	\\
F100& 5.4	& /	& /	\\
\\
V	& 1.2	& -3.2	& 0.99	\\
Vno	& 11	& /	& /	\\
M	& 20	& -3.8	& 0.091	\\
M-4 & 0.84	& -3.4	& 0.58	\\
M-10 & 9.1	& -3.7	& 0.13	\\
P	& 2.0	& -3.4	& 0.56	\\
P10	& 26	& -3.8	& 0.0	\\
P100& 1.6	& -3.4	& 0.53	\\

\hline
\end{tabular}
\tablecomments{
Results for pre-pollution of the IGM. 
Expected number, $N_{4.5}$, of stars with $\feoh < -4.5$ in the survey area of HES survey; 
metallicity, $\feoh_{\rm IGM}$, of intergalactic matter at z=0; 
and mass fraction, $\rm Fe_{PISN}/Fe_{IGM}$, of iron ejected by PISNe among total mass of iron ejected from mini-halos by SNe. }
\end{center}
\end{table}

\subsection{Negative Feedback by Dissociation Photon}

\begin{figure}
\plotone{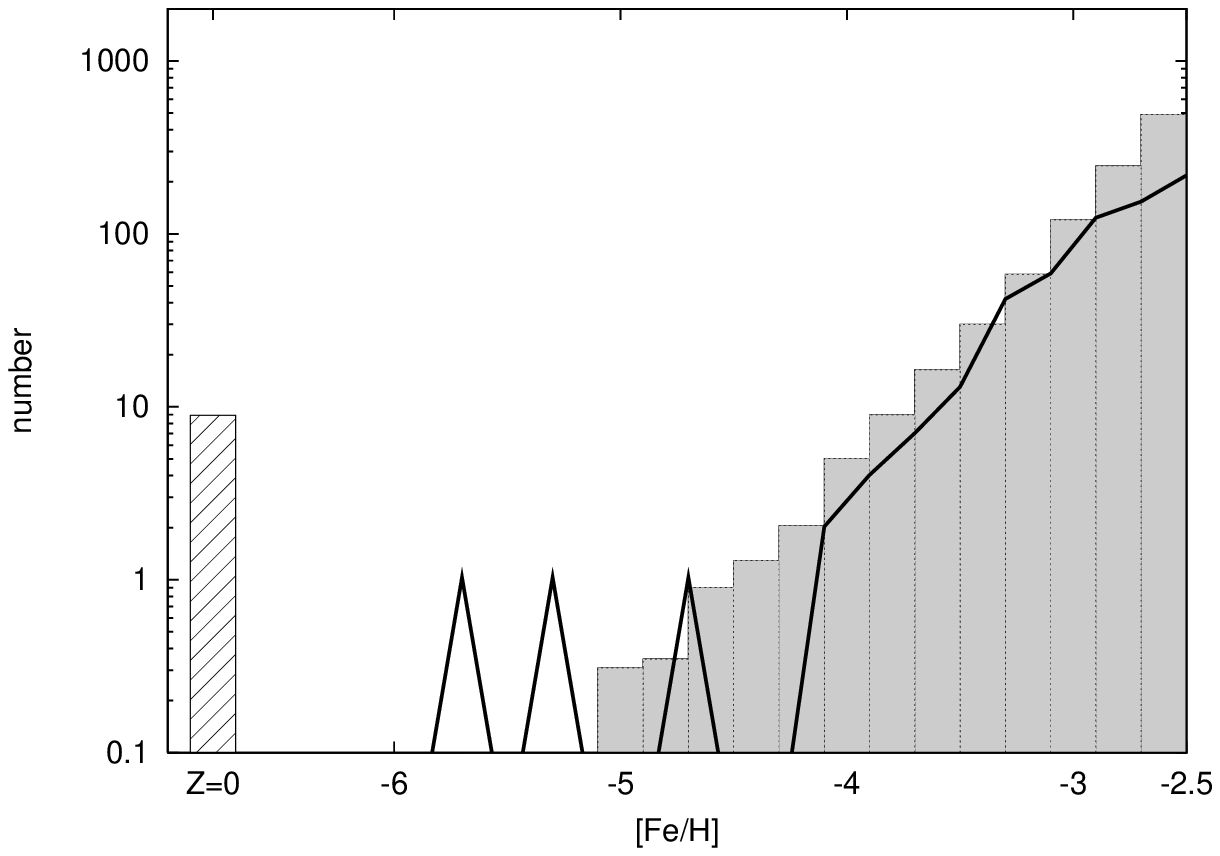}
\plotone{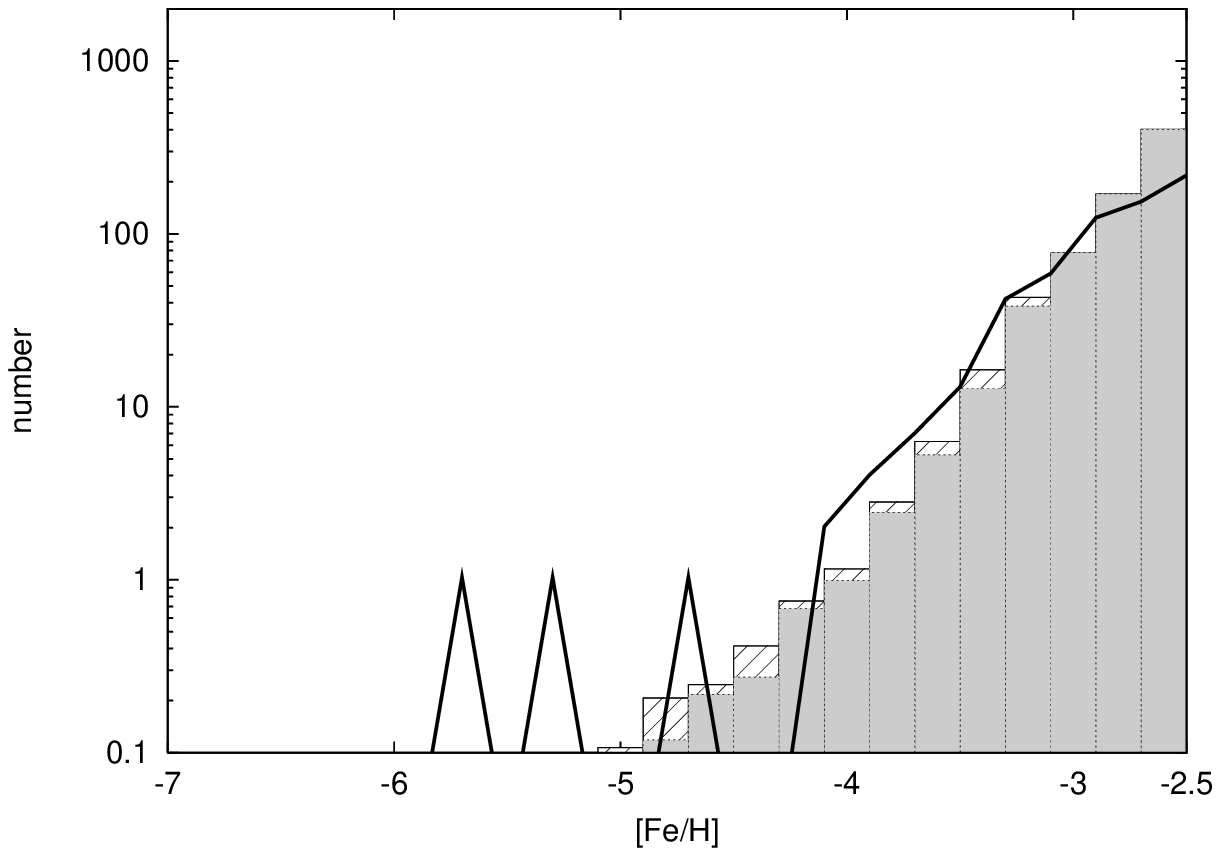}
\caption{Resultant MDFs for Models~Vno and V. 
   Hatched areas denote local first stars which formed before the first pollution in their host halo. 
   Shaded histograms show distribution of EMP stars formed after the first pollution. 
   Thick solid lines denote observed MDF. 
}
\label{V}
\end{figure}

Top and bottom panels of Figure~\ref{V} show the resultant MDFs for Models~Vno and V, i.e., the models with suppression of star formation by LW background without positive feedback. 
   Hatched areas in Fig.~\ref{V} and following figures in this section show the number of stars that formed before the first pollution in their host halo, and shaded histograms denote the MDF of stars formed after the first pollution. 
   We refer to these stars formed before the first pollution as local first stars in this paper. 
   All pristine iron in their interior comes from intergalactic matter polluted by energetic supernovae which blowout their host halo.

For these models, the minimum halo mass for star formation becomes larger after $z = 20$ because of negative feedback with LW background. 
   For model Vno, the number of Pop.~III survivors is much smaller than for Model~F0 or F-11 because the number of primordial mini-halos with star formation decreases with the increase in the minimum halo mass for star formation.  
   As shown in \S~\ref{accS}, these stars are possibly observed as HMP stars. 
   In such larger halos, SN ejecta suffers from the dilution with large mass and the metallicity of stars born after the first pollution becomes $\feoh \sim -5$. 
   Break of MDF at $\feoh\sim-4$ does not appear. 
   The predicted number of HMP stars is consistent with the observations but the assumption that gas is restricted in mini-halo even if PISN occurs is not realistic for the low-mass halos. 

For Model V, the predicted number of HMP stars results is too small because of the very massive IMF of primordial stars. 
   In our model, metallicity of IGM becomes $\feoh>-4.5$ at very high redshift ($z\sim20$) by ejecta of PISNe. 
   This result indicates that some positive feedback effect that enables low-mass star formation is required to form HMP stars. 
   At $z>20$, few Pop.~II stars are formed, and galactic wind originated from Pop.~II stars seems to be negligible for HMP stars.

\subsection{Metal cooling model}
\begin{figure}
\plotone{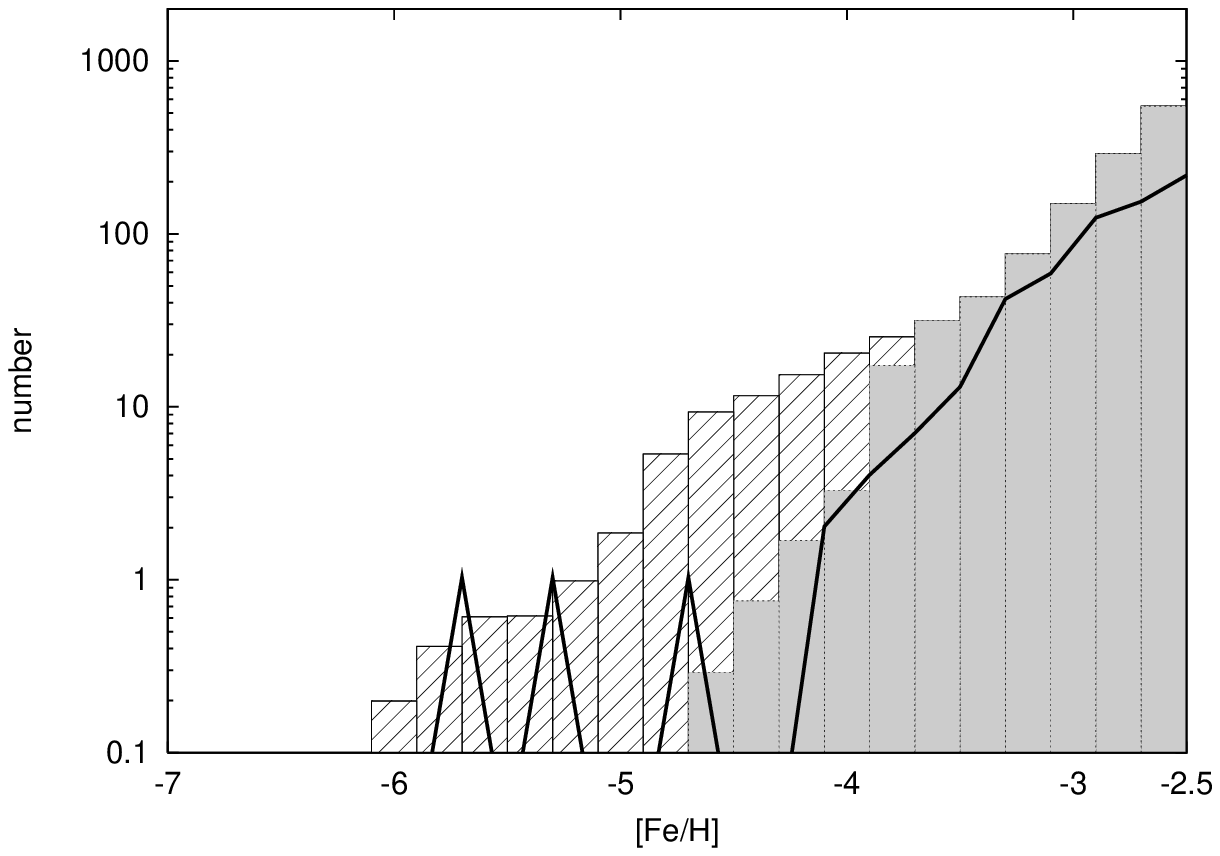}
\plotone{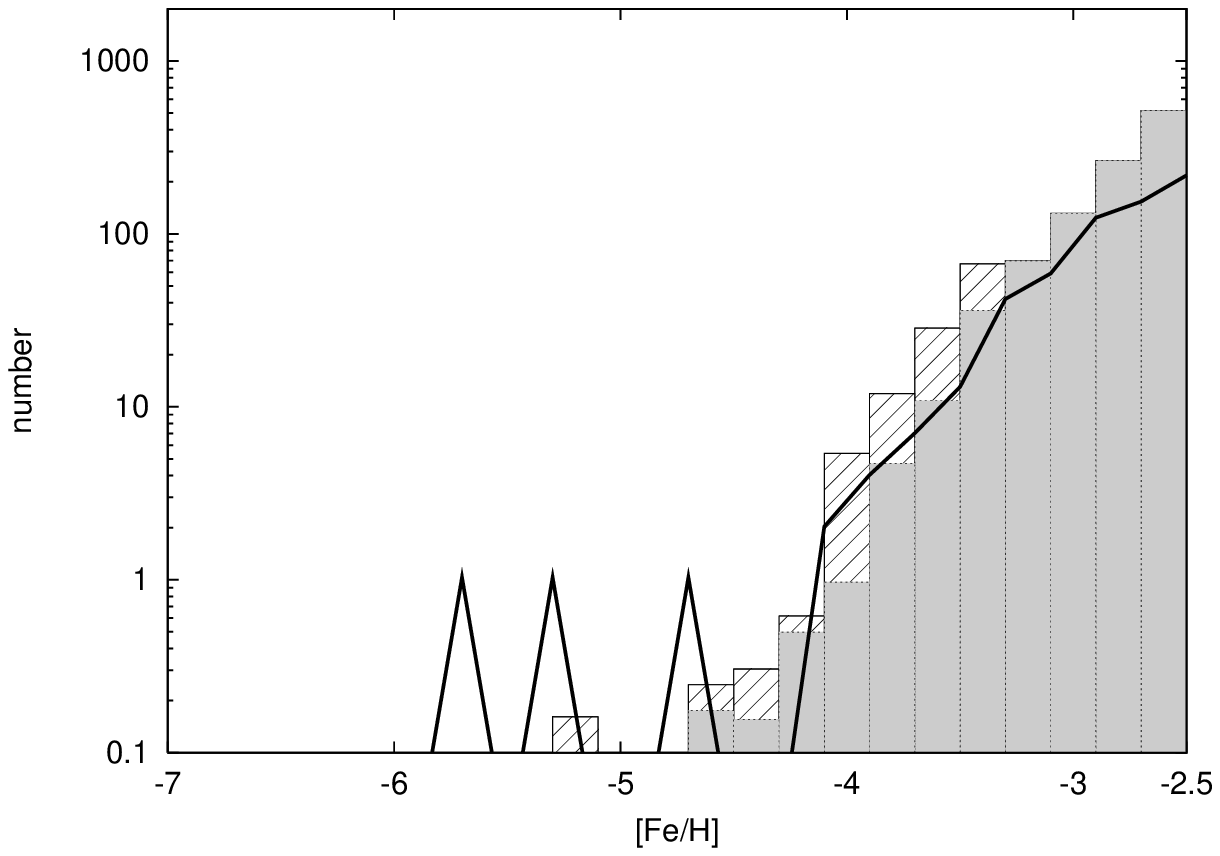}
\plotone{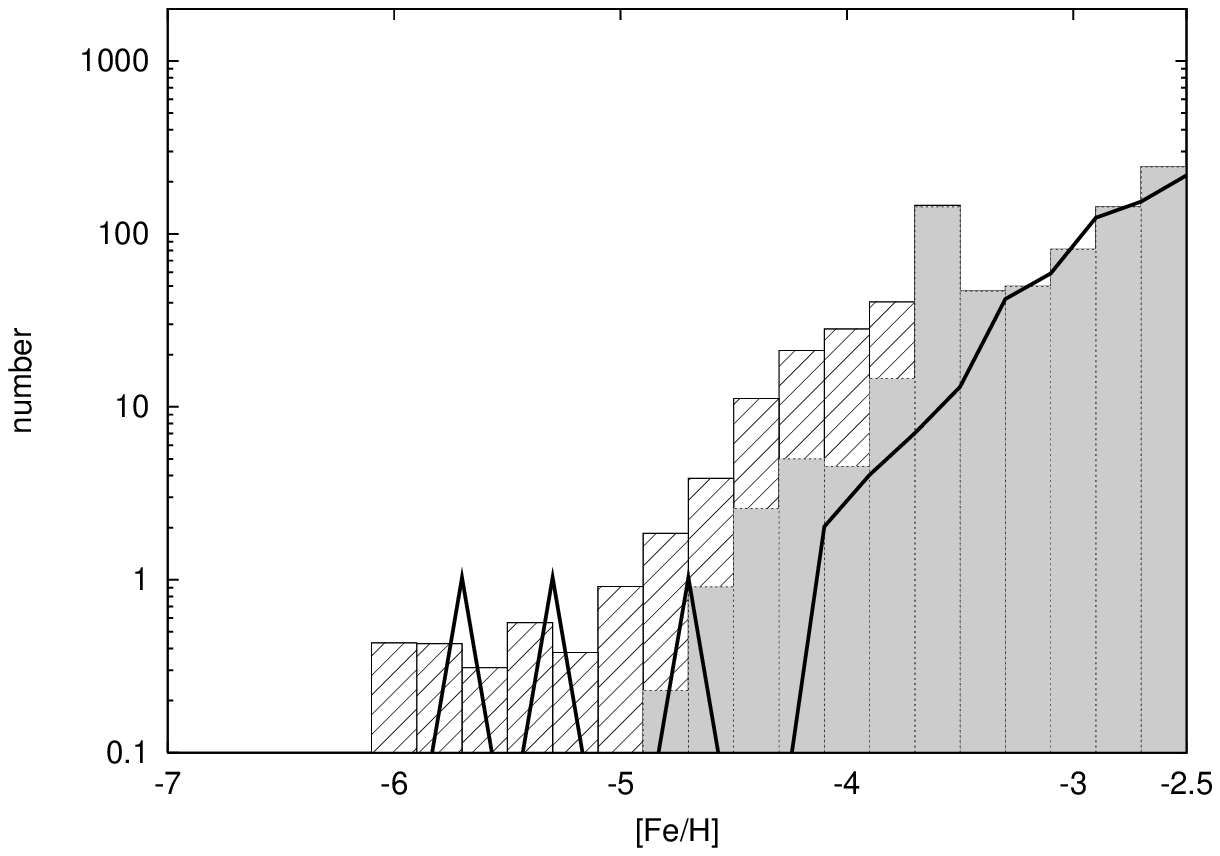}
\caption{Same as Figure~\ref{V} but for Models~M, M-4, and M-10
}
\label{M}
\end{figure}

The one possible catalyst for gas cooling and the formation of the second generation stars is metals. 
   Top, middle and bottom panels of Figure~\ref{M} show the resultant MDF of Models M, M-4, and M-10, respectively, with positive feedback by metals.  
   In these models, we assume that very massive stars are formed in the halos of the metallicity below $\feoh_{\rm cri}$, while the same high-mass IMF as for the EMP population is apply to the halos of larger metallicity.  
   We assume $\feoh_{\rm cri}=-6$ for Models M and M-11, and $\feoh_{\rm cri}=-4$ for Model M-4. 

The MDFs of stars formed after the first pollutions are very similar to model F and well resemble the observations. 
   For Models~M and M-10, stars with pre-pollution are distributed between $-6 \lesssim \feoh \lesssim -4$.  
   They predict 9 and 20 HMP stars, respectively. 
   This is comparable to observations if we take into account the fraction of stars with HDS follow-up and the heavy effect of surface pollution on main-sequence stars as discussed in \S~\ref{accS}. 
   For these models, most metals for pre-pollution are ejecta of Type~II SNe. 
   This is also consistent with the observations. 
   However, the number of local first stars with $-5 < \feoh < -4$ is larger as compared with the observations and the metallicity break disappears as a results of these stars.  
   If we consider the inhomogeneity in the metallicity distribution of IGM, however, the metallicity of these stars also spreads over a wider range, and this model can be compatible with the observations.  
   
In Model M-4, we can see MDF break at $[Fe/H]\sim-4$.
   But it predicts too small a number of HMP stars. 
   For this model, metals ejected by PISNe overwhelm metals ejected by Type~II SNe in IGM. 
   More than half of stars with $\feoh<-3.5$ are formed with the metal abundances of IGM, although no stars are identified with the abundance pattern characteristic of PISNe yields.  
   Although we cannot reject this model because of lack of statistics, Model~M seems to be more plausible. 

As shown in \S~\ref{sfeS}, there is a hump of MDF at $\feoh\sim-3.5$ for Model~M-10 with high SFE. 
   Results about pre-pollution are similar to Model~M and the metal pollution of IGM is almost independent of SFE, also.  
   This is because metal in the IGM is mostly provided by massive local first stars in mini-halos before the first pollution.  
   Only one massive star is formed before the first pollution in each halo because of local negative feedback, and after gas is blown off the halo, no stars are formed in the dark matter mini-halos until they accumulate a significant amount of gas again.  
   Therefore, the pre-pollution rate by primordial stars is weakly dependent on the SFE.  
   
\subsection{Photoionization model}
\begin{figure}
\plotone{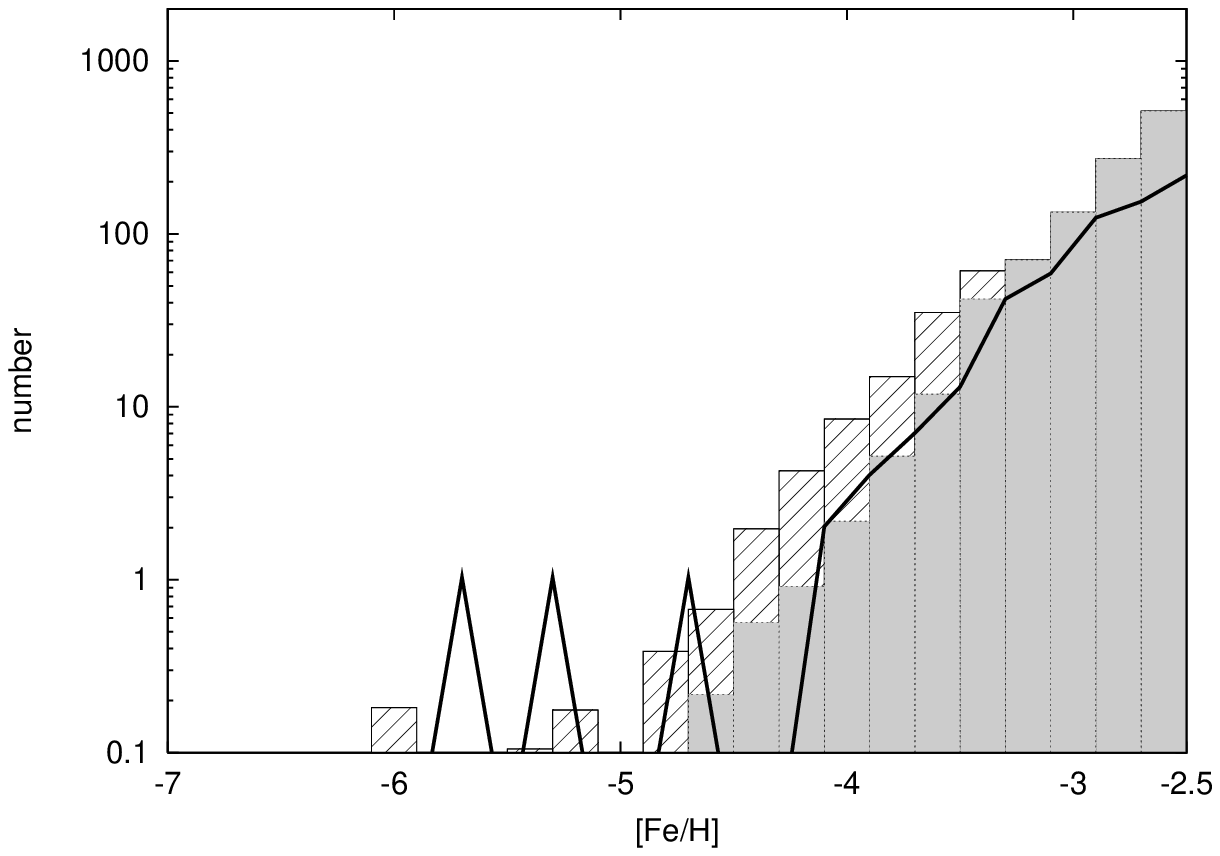}
\plotone{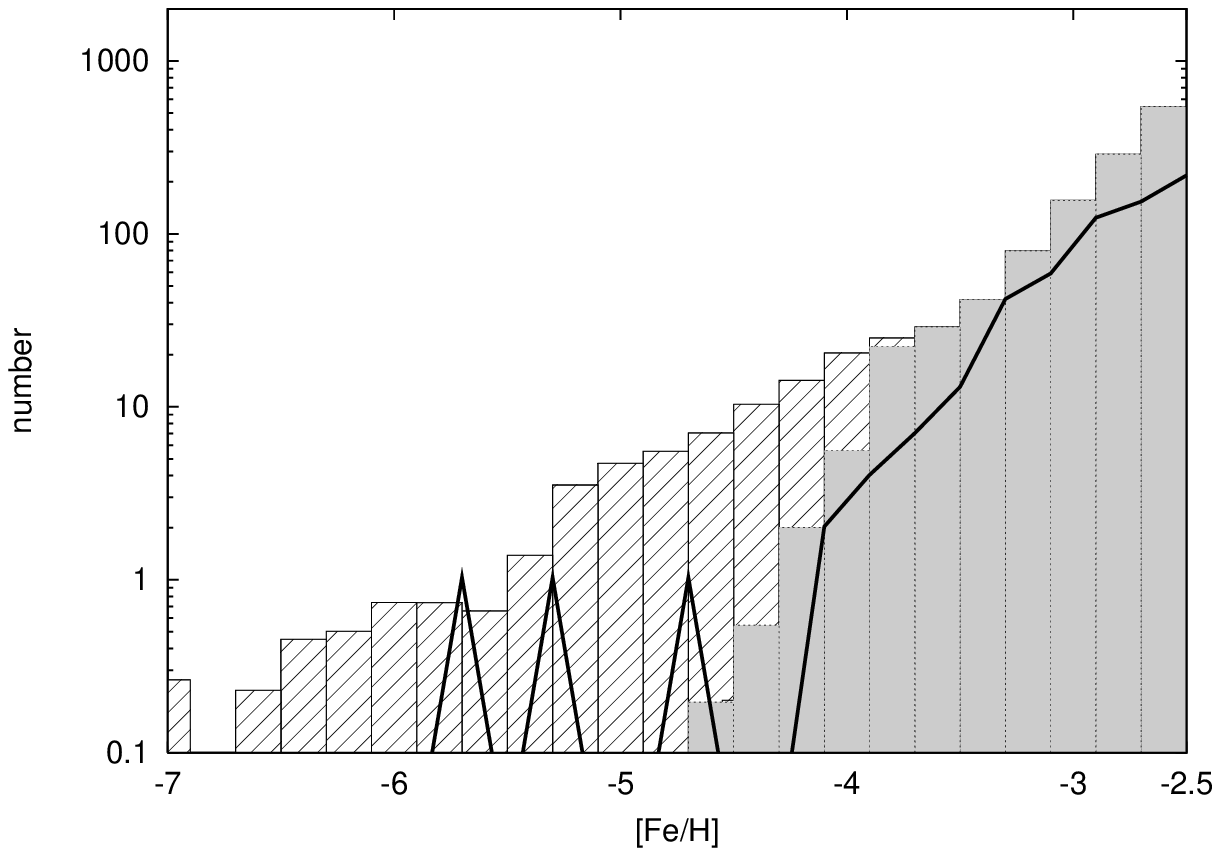}
\plotone{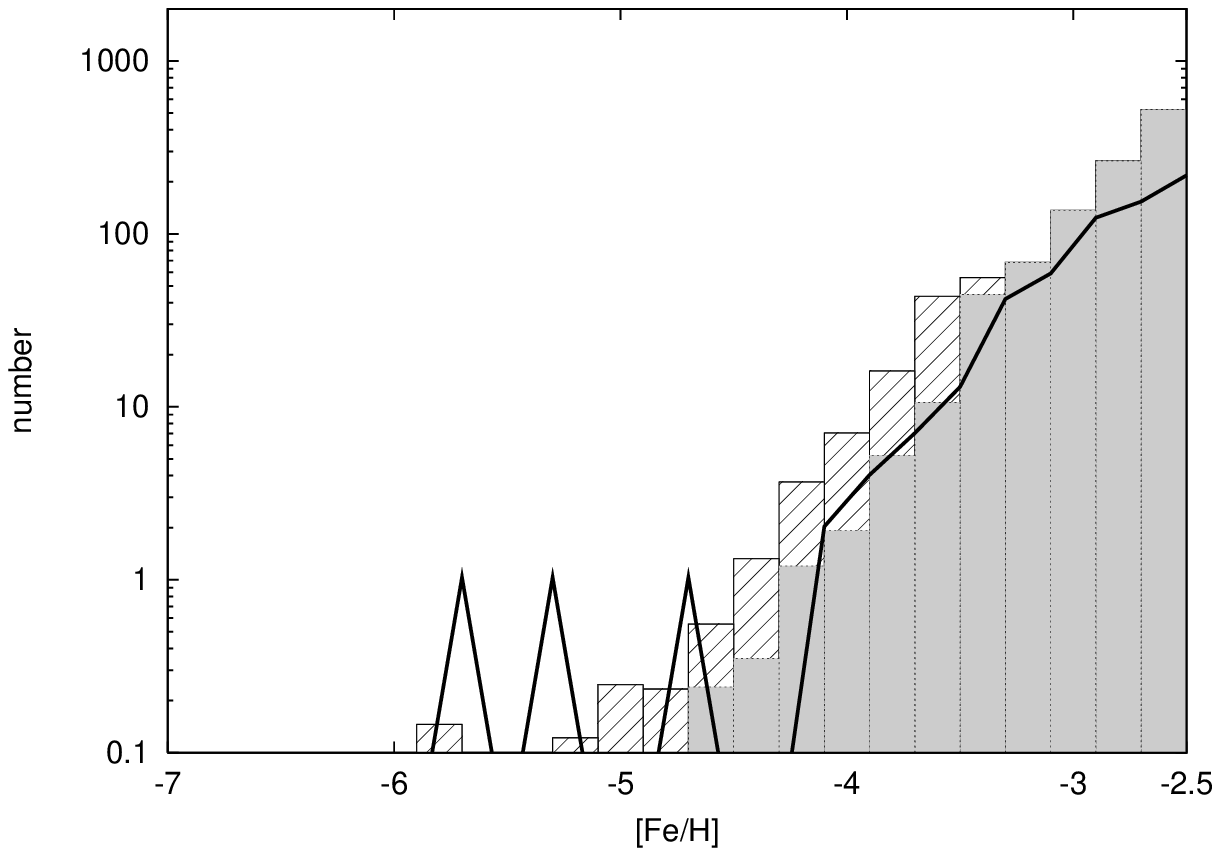}
\caption{Resultant MDFs of Models~P, P10 and P100.  
}
\label{P}
\end{figure}

The top panel of Figure~\ref{P} shows the resultant MDFs for Model~P with the photo-ionization positive feedback for the formation of the second generation stars.   
   The predicted number of HMP stars is slightly smaller than observed because of the pre-pollution. 
   For this model, MDF does not decrease so steeply at $\feoh \simeq -4$. 
   About a half of metals for IGM pre-pollution are provided by PISNe. 
   Stars, formed out of pre-polluted IGM before the first pollution, are distributed around $\feoh=-3.5- -5$, as shown by hatched histograms in the figure. 
   More than half of stars with $\feoh<-3.5$ are the local first stars. 
   The abundances of these stars should show the abundance pattern characteristic of PISNe yield but it should be blended with ejecta of Type II SNe.   
   
More observational studies of the MDF and the abundance patterns of stars with $\feoh<-3.5$ are necessary to test the validity of this model. 
   Further theoretical studies are also required for detailed metallicity distribution of the local first stars since our results indicate only typical metallicity of these stars by assuming homogeneous abundance for intergalactic matter.


\subsection{Typical mass of primordial stars} 

The predicted number of HMP stars depends on the assumption for IMF of primordial stars.   
   Middle and bottom panels of Figure~\ref{P} show the results of Models~P10 and P100. 
   For the Model~P10 with $M_{\rm md,p}=10 \msun$, many HMP stars are formed from the pre-polluted gas. 
   Because no stars are formed in the mini-halos of unionized gas under the influence of LW background, the predicted number of stars with $\feoh<-4$ becomes smaller than Model~F but is still much larger than the observation. 
   These results indicate that the fraction of low mass stars among primordial stars should be lower than for EMP stars. 

Despite stars of $200 \msun$ becoming PISNe and stars of $100\msun$ becoming black holes without metal ejection, the results of P with $M_{\rm md} = 200 \msun$ and P100 with $M_{\rm md} = 100 \msun$ are very similar. 
   In both models, a significant fraction of primordial stars become PISNe and a small number of HMP stars are formed.  
   This suggests the difficulty to know the specific value of typical mass and the mass distribution of primordial stars from the observations of EMP and HMP stars. 

\section{Discussion ---|HMP stars \& first low-mass stars}\label{discussion}

In the previous sections, we have shown that the first low-mass stars, formed before the first pollution, can have the surface iron abundance of $\feoh \sim -5$ through the ISM accretion and/or through the pre-pollution of energetic SNe in the very early universe to be observed as HMP stars. 
   In this section, we consider other properties of HMP stars and discuss the formation scenario of these stars.  

\subsection{Abundances of the light elements}\label{HMPabundance}

As for the HMP stars, one of the central issues is the origin of their peculiar abundance patterns \citep[e.g. see][]{Tumlinson07}, reported by HDS observations \citep{Christlieb02, Frebel05, Norris07, Bessell04, Aoki06}. 
   All the three HMP stars show large carbon enhancement with $\abra{C}{Fe}>+2$ and some stars show the enhancement of nitrogen and oxygen also. 
   The enrichments of Na, Ma and Al are also reported for two of them but with different extents.  

One plausible scenario for the abundances of these elements is the wind accretion of the envelope matter ejected from AGB companion in binary systems \citep{Suda04}. 
   In particular, lithium depletion of unevolved HMP star, HE1327-2326, indicates that its surface is covered by the matter processed by the evolved stars \citep{Aoki06, Tumlinson07}. 
   \citet{Nishimura09} investigate the nucleosynthesis in the intermediate-mass stars with $Z=0$ during TP-AGB phase and show that the abundance patterns of HMP stars can be explained by accretion from AGB companion, including the differences among three HMP stars. 
   Surface carbon abundances of the main sequence secondary stars are changed to $\abra{C}{H}>-3$ by the wind accretion in the binary with separation smaller than $\lesssim 1000$AU \citep{Komiya07}

\citet{Maynet06} argue that stellar wind from massive Pop.~III star with rapid rotation brings about abundance pattern similar to that of HMP stars.  
   For these massive primaries, however, the fast wind velocity prevents the low-mass secondary to accrete sufficient wind matter.  
   The scenario with peculiar SN yield, proposed by \citet{Iwamoto05}, has the advantage of letting us discuss the peculiar abundance pattern for the light elements to the iron group elements in one lump, but difficulties have been pointed out about the lithium depletion in HE1327-2326 as well as about the formation of stars with high carbon abundance with the limited amount of carbon ejected \citep[see e.g.,][]{Nishimura09}.  


\subsection{Number of Pop.~III and EMP survivors}

Observationally, three HMP stars are detected among the sample stars of HES surveys and several more stars are expected from the frequency of follow-up with high dispersion observations.  
   On the other hand, we predict 300 stars should be in survey volume in HES survey in the fiducial model and 60 stars in the models with local radiative feedback.  
   As seen in Figure~\ref{accretion}, they disguise their surface with the accreted ISM of abundance $\feoh > -4$ during the main sequence stars and possibly are lost among the EMP stars.  
   Then, the number of stars with surface metallicity $\feoh<-4$ shrinks by about half. 
   For model with cosmological parameter by WMAP 5-year, the predicted number of Pop.~III stars decreases by $\sim10\%$, and yet, the predicted number is larger than that expected from the observations. 
   This means that the formation rate of low-mass secondary stars in the binary systems is lower for the primordial stars.  
   The number of Pop.~III survivors predicted by some models with $\mmd \gtrsim 100\msun$ for Pop.~III stars reduces to be comparable with observations. 
   In conclusion, the IMF is of higher-mass and/or the low-mass star formation is less efficient in the binary systems before the first pollution in the mini-halos.  

Our semi-analytic model lacks information on the spatial distribution of stars. 
   If there is a significant spatial metallicity gradient for EMP stars in the Galactic halo, it makes the comparison with the observed MDF not so reliable.  
   The spatial distribution of EMP stars yet remains unrevealed.
   Recently, \citet{Carollo07} report that the metallicity of outer-halo stars is lower than that of inner-halo stars, but most of their sample stars have metallicity of $\feoh>-2.5$. 
   From our results, we rather expect only weak radial gradient of metallicity at least for EMP stars since there is little difference in the formation epoch between the first low-mass survivors and EMP survivors as they are formed in many mini-halos almost independently of the redshift, as shown in \S~\ref{history} and Fig.~\ref{zFe}.  
   For the stars with $\feoh >-2.5$ in the Galactic halo, on the other hand, \citet{Ryan91} derive halo MDF with a peak around $\feoh\sim -1.8$ for kinematically selected sample stars. 
   In our fiducial model, the metallicity of low-mass survivors formed at $z=10$ reaches $\feoh=-1.5 - -2$.  
   This indicates that the formation of stars in the nearby Galactic halo has been suppressed after $z\sim10$. 
   Residual gas may go to form the Galactic bulge and/or the Galactic disk components through the merging processes.  

\subsection{Formation of the HMP stars}

Our results give insights into the formation of the first low-mass stars.  
   The abundance patterns of HMP stars show that all are the secondary members of binaries with moderate separation and probably with low- and intermediate-mass stars.  
   The HMP stars, known to date, all exhibit large carbon enrichments, which is possible only for the close binaries of separations $ \sim 10 - 100$ AU \citep{Suda04,Komiya07}.   
   This suggests that the low-mass stars can be formed exclusively in the close binary before the first pollution. 

From a theoretical perspective, it is known that the temperature and Jeans mass should be higher for the primordial gas clouds than for the star forming clouds with metals, because of the absence of the effective coolants such as metals and dust.  
   However, when the gas density becomes higher than $\sim 10^{10} \hbox{ cm}^{-3}$, three-body reactions work and promote the formation of the H$_2$ molecule to lower the Jeans mass \citep{Palla83}. 
   It is argued that H$_2$ formation on the dust surface also works to change thermal evolution from the primordial cloud only in the phase of high density phase $\sim 10^{10} \hbox{ cm}^{-3}$ \citep{Omukai05}. 
   The radial extension of such high-density regions is $\sim 10^{3}$ AU.  
   Before the first pollution, therefore, low-mass fragmentation can occur only in the contracting gas clouds, which may lead to the formation of low-mass survivors in close binary systems.  
   Recently, \citet{Machida08} investigate the evolution of first star forming core and showed that fragmentation may occur in a later collapsing phase.  
   \citet{Clark08} also demonstrate the formation of ``first stellar cluster" by fragmentation at high density.  
   More investigations are required to reveal the conditions for the formation of the first low mass stars, and yet, our results suggest that low-mass stars can be formed only in high density cloud cores as secondary members of binary systems with lower efficiency than EMP stars.  

\section{Conclusions}\label{summary}

We have studied star formation and chemical evolution in the early universe from the perspective of the hierarchical structure formation of the Galaxy by constructing the merging trees on the $\Lambda$ cold dark matter scenario according to the extended Press-Schechter theory. 
   In particular, we investigate the formation history of EMP and HMP stars and the origin of iron in HMP stars.  
   We demonstrate that the hierarchical chemical evolution can reproduce the characteristics of MDF and the frequency of EMP stars in the Galactic halo by applying the high-mass initial mass function and the contribution of binary, derived for EMP stars in our previous work \citep{Komiya07,Komiya09}.   

Our main conclusions are summarized as follows.  

\begin{enumerate}
\item 
   In the hierarchical structure formation scenario, the metal-poor stars are divided into two groups, according to whether they are born before or after the first pollution, i.e., the first SNe pollution of the gas in each mini-halo.  
   The stars of the former group are made of gas with the pristine abundances of mini-halos, and the stars of the latter group are enriched with metals, produced by their own SNe of each halo.  
   The latter stars well reproduce the observed metallicity distribution of EMP stars of $\feoh \gtrsim -4$ and the observational counterparts of the former stars are found in the HMP stars of $\feoh < -4$.  

\item 
   While star formation efficiency has few effects except for the stars of the lowest metallicity, the initial mass function directly influences the frequencies of low-mass stars, survived to date for given metallicity.  
   For EMP stars of $\feoh \gtrsim -4$, the high-mass IMF of medium mass $\sim 10 \msun$ with the contribution of binaries, derived on the basis of the statistics of CEMP and EMP stars by \citet{Komiya07, Komiya09}, gives the total number of EMP stars, compatible with the observations by HES surveys.   
   In particular, the low-mass stars observed in the Galactic halo today are mostly (more than 90\%) born as the secondary members of binaries.  
   On the other hand, the low-mass IMF such as derived for the Galactic spheroid component gives rise to the overabundance of EMP survivors by more than two orders of magnitude. 

\item  
   For stars born before the first pollution, the scarcity of HMP stars suggests that the star formation, especially of low-mass survivors, is less efficient as compared with that for the EMP population of $-4 \lesssim \feoh \lesssim -2.5$.  
   The overproduction of low-mass survivors still persists by a factor of $\sim 10$ even if we take into account the suppression of the star formation efficiency due to the local radiation feedback of the first massive stars.   
   This is indicative of higher-mass IMF and/or less efficient binary formation in the gas of the lowest metallicity than for the EMP populations.  

\item  
   As for higher metallicity of $\feoh \gtrsim -2.5$, there is no indication of significant changes in the IMF, in particular into a low-mass IMF as long as the fields stars in the Galactic halo are concerned.   
   From the comparison with the MDF derived by \citet{Ryan91} for kinematically selected samples, however, the star formation efficiency declines for $\feoh \gtrsim -1.8$.  
   This occurs near the redshift $z \sim 10$ in our fiducial model and may be related to the formation of the Galactic bulge and/or the Galactic discs.

\item 
   For the first low-mass stars, we have evaluated the effect of ISM accretion after birth by taking into account the dynamical and chemical evolution of the Galaxy.  
   The accretion in the mini-halos, in which Pop.~III stars are formed, dominates the surface pollution, giving a much larger amount of accreted gas and metals than in the evolved massive halos.   
   If we take into account the cooling of gas and resultant concentration of gas and stars into the central region of their birth halos, the ISM accretion can give rise to the surface pollution sufficient to explain the iron abundances observed for HMP stars.  

\item 
   We have investigated the feedback effect from primordial stars, born prior to EMP stars, and the pre-pollution of intergalactic gas by SNe in the first collapsed halos.  
   Primordial supernovae pollute intergalactic matter to $\feoh=-3.2 - -3.8$ in our models and some EMP survivors with such metallicity can be formed with pre-pollution before the first pollution. 
   Metal ejecta from Type II SNe that occurred in the mini-halos with positive feedback also pollute intergalactic matter. 
In the models with significant feedback effects, the ejecta from PISN with a characteristic abundance pattern should blend and can be consistent with observational absence of abundance pattern peculiar to PISNe.  
   Detection of 3 HMP stars indicates that some positive feedback enables the low mass star formation below $\feoh<-4$. 
\end{enumerate}

From the binary scenario, we have revealed that there are two possible origins of iron for HMP stars, the surface pollution of accreted IMF in the mini-halos and the pre-pollution of intergalactic gas with metals that energetic SNe have blown off the host halos.    
   The polluted Pop.~III stars with ISM accretion and the pre-polluted HMP stars differ only in the presence of pristine iron in their interior. 
   In order to distinguish these two possibilities, further investigations are necessary to reveal the effects that the pristine metals of such small abundances as corresponding to the iron abundance observed for HMP stars have on the nucleosynthesis and evolution of low-mass and intermediate-mass stars \citep[e.g., see][]{Suda09b}.  

In this paper, we use the simplified assumption about the star formation and deal only with iron.  
   In particular, we assume the instantaneous and homogeneous mixing of SN ejecta in the halos and in the IGM.  
   In actuality, however, we should take into account the dependence of the expansion of shells on the explosion energy of SNe and host halo mass, and the finite timescale of mixing \citep{Greif08, Tornatore07}.  
   It little affects the present results as long as the host halos are not disrupted by SNe, however, since the SNe shells involve the circumstellar matter comparable to the smallest masses in the mini-halos that can nurture the star formations, except that it may reduce the number of stars with the smallest metallicity of EMP stars, which are formed in the halos of larger masses that collapse later.   
   From our homogeneous mixing model, we cannot compute detailed MDF of stars formed before the first pollution and additional investigation is required to distinguish the origin of positive feedback. 
   Further elaborations are necessary for the interactions between the star formation and SNe and the binary formation under the metal-deficient condition.  
   We may expect to gain a better understanding of EMP stars and the evolution of the Galaxy from the large-scaled, deep surveys of halo stars now in progress such as SDSS/SEGUE and LAMOST, as well as from the detailed abundance studies with large telescopes.  

This work is supported partly by JSPS Grand-in-aid for Scientific Research (18104003, 18072001, 19740098).  
A part of results are reported in AIP~Conference~990 "First Stars", AIP~Conference~1016 "The Origin of Matter and Evolution of Galaxy", and IAU~Symposium~255 "Low-Metallicity Star Formation: from First Stars to Dwarf Galaxies".


\begin{thebibliography}{}
\bibitem[Abel et al.(2002)]{Abel02} Abel, T., Bryan, G. L., \& Norman, M. L. \ 2002, Science, 295, 93
\bibitem[Aoki et al.(2006)]{Aoki06} Aoki, W., et al. 2006, \apj, 639, 897
\bibitem[Aoki et al.(2007)]{Aoki07} Aoki, W., Beers, T.C., Christlieb, N., Norris, J.E., Ryan, S.~G., \& Tsangarides, S., 2007 \apj, 655, 492
\bibitem[Barkana \& Loeb(2001)]{Barkana01} Barkana, R., \& Loeb, A.\ 2001, \physrep, 349, 125
\bibitem[Bate \& Bonnel(1997)]{Bate97} Bate, M.R. \& Bonnel, I.A. 1997, \mnras, 285, 33
\bibitem[Beers et al.(1992)]{Beers92} Beers, T.C., Preston, G.W., \& Shectman, S.A. 1992, \aj, 103, 1987
\bibitem[Beers \& Christlieb(2005)]{Beers05a} Beers, T.C. \& Christlieb, N. 2005, ARA\&A, 43, 451
\bibitem[Beers et al.(2005)]{Beers05} Beers, T.C., Christlieb, N., Norris, J.E., et al.
2005, IAUS, 228, 175
\bibitem[Bessell et al.(2004)]{Bessell04} Bessell, M. S., Christlieb, N., \& Gustafsson, B. 2004, \apjl, 612, 61
\bibitem[Bond et al.(1991)]{Bond91} Bond, J.R., Cole, S., Efstathiou, G., et al. 
1991, \apj, 379, 440 
\bibitem[Bromm \& Loeb(2003)]{BrommL03} Bromm, V., \& Loeb, A., 2003, Nature, 425, 812
\bibitem[Bromm \& Larson(2004)]{Bromm04} Bromm, V., \& Larson, R.~B., 2004, ARA\&A, 42, 79
\bibitem[Bromm et al.(1999)]{Bromm99} Bromm, V, Coppi, P. S., Larson, R. B., 1999 ,\apjl, 527, L5
\bibitem[Bromm et al.(2003)]{Bromm03} Bromm, V., Yoshida, N., \& Hernquist, L., 2003, \apjl 586, L135
\bibitem[Carollo et al.(2007)]{Carollo07} Carollo, D., Beers, T.~C., et al. 2007, Nature, 450, 1020
\bibitem[Chabrier(2003)]{Chabrier03} Chabrier, G.\ 2003, \pasp, 115, 763 
\bibitem[Christlieb et al.(2001)]{Christlieb01} Christlieb, N., Green, P.J., et al.
2001, \aap, 375, 366
\bibitem[Christlieb et al.(2002)]{Christlieb02} Christlieb, N., Bessell, et al.
2002 Nature, 419, 904-906
\bibitem[Christlieb(2003)]{Christlieb03} Christlieb, N.\ 2003, Rev Mod. Astron,16, 191
\bibitem[Christlieb et al.(2008)]{Christlieb08} Christlieb, N., Sch{\"o}rch, T. Frebel, A., et al.
2008, \aap, 484, 732
\bibitem[Clark et al.(2008)]{Clark08} Clark, P. C., Glover, S. C. O., Klessen, R. S., 2008, \apj, 672, 757	
\bibitem[Frebel et al.(2005)]{Frebel05} Frebel, A., Aoki, W., Christlieb, N., et al. 2005 Nature 434, 871-873
\bibitem[Frebel et al.(2009)]{Frebel09} Frebel, A., Johnson, J. L., Bromm, V., 2009, \mnras, 392, L50 
\bibitem[Fujimoto et al.(1990)]{Fujimoto90} Fujimoto, M.~Y., Iben, I.~Jr., \& Hollowell, D.\ 1990, \apj, 349, 580 
\bibitem[Fujimoto et al.(1995)]{Fujimoto95} Fujimoto, M.Y., Sugiyama, K., Iben, I. Jr., \& Hollowell, D. 1995, \apj , 444, 175 
\bibitem[Fujimoto et al.(2000)]{Fujimoto00} Fujimoto, M.~Y., Ikeda, Y., \& Iben, I.~Jr.  2000, \apj, 529, L25
\bibitem[Greif et al.(2008)]{Greif08} Greif, T.~H., Johnson, J.~L., Bromm, V., Klessen, R.~S. 2007, \apj, 670, 1
\bibitem[Heger \& Woosley(2002)]{Heger02} Heger, A., \& Woosley, S. E. 2002, \apj, 567, 532
\bibitem[Helmi et al. (2006)]{Helmi06} Helmi, A., Irwin,M.~J., Tolstoy, E., Battaglia, G., Hill, V., Jablonka, P., Venn, K., Shetrone, M., Letarte, B., Arimoto, N., and 6 coauthors 2006, \apj, 651L, 121 
\bibitem[Honda et al.(2004)]{Honda04}  Honda, S., Aoki, W., Kajino, T., Ando, H., Beers, T.~C., Izumiura, H., Sadakane, K., \& Takada-Hidai, M.\ 2004, \apj, 607, 474 
\bibitem[Iben(1983)]{Iben83} Iben, I., 1983, Mem. S. A. It., 54, 321
\bibitem[Iwamoto et al.(2005)]{Iwamoto05}Iwamoto, N., Umeda, H., et al. 
2005, Science, 309, 15
\bibitem[Karlsson(2005)]{Karlsson05} Karlsson, T.  2005, \aap, 439, 93 
\bibitem[Karlsson(2006)]{Karlsson06} Karlsson T.  2006, \apjl, 641, L41
\bibitem[Kirby et al.(2008)]{Kirby08} Kirby E.~N, Simon, J.~D., Geha, M., Guhathakurta, P. and Frebel, A. 2008, \apjl, 43, 46
\bibitem[Kitayama \& Yoshida(2005)]{Kitayama05} Kitayama, T., \& Yoshida, N. \apj, 630, 675
\bibitem[Komiya et al.(2007)]{Komiya07} Komiya, Y., Suda, T., Minaguchi, H., Shigeyama, T., Aoki, W., \& Fujimoto, Y. M. 2007 \apj 658, 367 (Paper~I)
\bibitem[Komiya et al.(2009a)]{Komiya09} Komiya, Y., Suda, T., \& Fujimoto, Y. M. 2009, \apj, 694, 1577 (Paper~II)
\bibitem[Komiya et al.(2009b)]{Komiya09L} Komiya, Y, Suda, T., \& Fujimoto, Y. M. 2009, \apjl, 696, L79 (Paper~III)
\bibitem[Lacey \& Cole(1993)]{LC93} Lacey, C., \& Cole, S.\ 1993, \mnras, 262, 627 
\bibitem[Limongi et al.(2003)]{Limongi03} Limongi, M., Chieffi, A. \& Bonifacio, P, \apj, 594,L123
\bibitem[Machida et al.(2005)]{Machida05} Machida, M.N., Tomisaka, K., et al. Nakamura, F., \& Fujimoto, M.Y. 2005, \apj, 622, 39
\bibitem[Machida et al.(2008)]{Machida08} Machida, M.N., Omukai, K., Matsumoto, T., \& Inutsuka, S.2008, \apj, 677, 813
\bibitem[Maynet et al.(2006)]{Maynet06} Maynet, G., Ekstr{o}m, S., \& Maeder, A., \aap, 447, 623
\bibitem[Nakamura \& Umemura(2001)]{Nakamura01} Nakamura, F., \& Umemura, M., \apj, 548, 19
\bibitem[Nishi \& Susa(1999)]{Nishi99} Nishi, R., \& Susa, H. 1999, \apj, 523, L103 
\bibitem[Nishimura et al.(2009)]{Nishimura09} Nishimura, T, Suda, T. Aikawa, M. et al. 2008, submitted to PASJ
\bibitem[Norris et al.(2007)]{Norris07} Norris, J.E., Christlieb, N., Korn, A. J.,  et al.
2007 \apj, 670, 774
\bibitem[Ochi et al. (2005)]{Ochi05}Ochi, Y. Sugimoto, K. \& Hanawa, T. 2005, \apj, 623, 922
\bibitem[Omukai \& Nishi (1999)]{Omukai99} Omukai, K. \& Nishi, R. 1999, \apj, 518, 64
\bibitem[Omukai \& Para (2003)] {Omukai03} Omukai, K. \& Para, F. 2003, \apj, 589, 677
\bibitem[Omukai et al.(2005)]{Omukai05} Omukai, K., Thuribe, T., Schnerder, R., \& Ferrara, A., 2005, \apj, 626, 627
\bibitem[O'Shea \& Norman (2007)]{O'Shea07} O'Shea, B.~W., \& Norman, M.~L., 2007, \apj, 654, 66
\bibitem[Palla et al.(1983)]{Palla83} Palla, F., Salpeter, E. E. \& Stahler, S. W. 1983, \apj,  271, 632
\bibitem[Prantzos(2003)]{Prantzos03} Prantzos, N. 2003, \aap, 404, 211
\bibitem[Press \& Schechter(1974)]{PS74} Press, W.~H., \& Schechter, P.\ 1974, \apj, 187, 425
\bibitem[Ricotti et al.(2002)]{Ricotti02} Ricotti, M., Gnede, N.~Y., \& Shull, J.~M. 2002, \apj, 575, 49 
\bibitem[Rossi et al.(1999)]{Rossi99} Rossi, S.~C.~F., Beers, T.~C., \& Sneden, C.\ 1999, ASP Conf.~Ser.~165: Stromlo Workshop on the Galactic Halo, 165, 264
\bibitem[Ryan \& Norris(1991)]{Ryan91} Ryan, S.G. \& Norris, J.E. 1991, \aj, 101, 1865
\bibitem[Salvadori et al.(2006)]{Salvadori06} Salvadori, S., Schneider, R., \& Ferrara, A. 2006, \mnras, 381, 647
\bibitem[Schaerer(2002)]{Schaerer02} Schaerer, D. 2002,\aap, 382, 28
\bibitem[Schneider(2006)]{Schneider06} Schneider, R., Omukai, K., Inoue, A.~K., \& Ferrara, A. 2006, \mnras, 369, 1437
\bibitem[Sch\"{o}erck et al. (2008)]{Schorck08} Sch\"{o}erck, T.,  Christlieb, N., et al. 2008, arXiv, 0809. 1172
\bibitem[Searle \& Zinn(1978)]{Searle78} Searle, L., \& Zinn, R. 1978, \apj, 225, 357
\bibitem[Somerville \& Kolatt(1999)]{SK99} Somerville, R.~S., \& Kolatt, T.~S.\ 1999, \mnras, 305, 1 
\bibitem[Stacy et al.(2009)]{Stacy09} Stacy, A., Greif, T. H.,\& Bromm, V. 2009, arXiv:0908.0712
\bibitem[Suda et al.(2004)]{Suda04} Suda, T., Aikawa, M., Machida, et al.
\bibitem[Suda et al.(2008)]{Suda08} Suda, T., Katsuta, Y., Yamada, S., et al. 2008, PASJ, 60, 1159
\bibitem[Suda et al.(2009)]{Suda09} Suda, T., Katsuta, Y., Yamada, S., et al. 2009, in prep. 
\bibitem[Suda \& Fujimoto(2009)]{Suda09b} Suda, T., \& Fujimoto, M.~Y., 2009, submitted to \apj
\bibitem[Tan \& McKee(2004)]{Tan04} Tan, J.~C., \& McKee, C.~F. 2004, \apj, 603, 383
\bibitem[Tegmark et al.(1997)]{Tegmark97} Tegmark, M., Silk, J., Rees, et al.
\bibitem[Tornatore et al.(2007)]{Tornatore07} Tornatore, L., Ferrara, A., \& Schneider, R. 2007, \mnras, 382,945
\bibitem[Trenti \& Shull (2010)]{Trenti10} Trenti, M., Shull, J.~M., \apj, 712, 435
\bibitem[Tumlinson(2006)]{Tumlinson06} Tumlinson, J. 2006, \apj, 641, 1
\bibitem[Tumlinson(2007b)]{Tumlinson07} Tumlinson, J. 2007, \apj, 665, 1361
\bibitem[Tumlinson(2007a)]{Tumlinson07L} Tumlinson, J. 2007, \apjl, 664, 63
\bibitem[Turk et al.(2009)]{Turk09} Turk, M.~J., Abel, T., \& O'shea, B. 2009, Science, 325, 601
\bibitem[Uehara \& Inutsuka(2000)]{Uehara00} Uehara, h. \& Inutsuka, S., 2000, \apj, 465, 608
\bibitem[Umeda \& Nomoto(2003)]{Umeda03} Umeda, H., \& Nomoto, K. 2003, Nature, 422, 871
\bibitem[Umeda \& Nomoto(2002)]{Umeda02} Umeda, H., \& Nomoto, K. 2002, \apj, 565, 385
\bibitem[Yoshida et al.(2008a)]{Yoshida08a} Yoshida, N., Oh, S. P., Kitayama, T., \& Hernquist, L., 2008, \apj, 663, 687
\bibitem[Yoshida et al.(2008b)]{Yoshida08}Yoshida, N., Omukai, K., \& Hernquist, L. 2008, Science, 321, 669
\bibitem[Yoshii(1981)]{Yoshii81} Yoshii, Y. \aap, 97, 280
\end{thebibliography}
\end{document}